%% file: v1_arXiv_silver_nitrides_article.tex
\documentclass[%
reprint,
superscriptaddress,
 amsmath,amssymb,
 aps,
prb,
floatfix,
]{revtex4-1}

\usepackage{graphicx}
\usepackage{dcolumn}
\usepackage{bm}
\usepackage[bookmarks=true,colorlinks=true,linkcolor={blue},citecolor=blue,urlcolor=blue]{hyperref} 
\usepackage[mathlines]{lineno}

\usepackage{longtable}	
\usepackage{tabularx}	
\usepackage{array}		
\usepackage{multirow}	
\usepackage{array}		
\usepackage{longtable}
\begin{document}


\title{Theoretical calculations on the structural, electronic and optical properties of bulk silver nitrides}


\author{Mohammed S. H. Suleiman}
\email[Corresponding author: ]{suleiman@aims.ac.za}
\affiliation{School of Physics, University of the Witwatersrand, Johannesburg, South Africa.}
\affiliation{Department of Physics, Sudan University of Science and Technology, Khartoum, Sudan.}

\author{Daniel P. Joubert}
\homepage[Homepage: ]{http://www.wits.ac.za/staff/daniel.joubert2.htm}
\affiliation{School of Physics, University of the Witwatersrand, Johannesburg, South Africa.}

\date{\today}
\begin{abstract}
We present a first-principles investigation of structural, electronic and optical properties of bulk crystalline Ag$_3$N, AgN and AgN$_2$ based on density functional theory (DFT) and many-body perturbation theory. The equation of state (EOS), energy-optimized geometries, cohesive and formation energies, and bulk modulus and its pressure derivative of these three stoichiometries in a set of twenty different structures have been studied. Band diagrams and total and orbital-resolved density of states (DOS) of the most stable phases have been carefully examined. Within the random-phase approximation (RPA) to the dielectric tensor, the single-particle spectra of the quasi electrons and quasi holes were obtained via the GW approximation to the self-energy operator, and optical spectra were calculated. The results obtained were compared with experiment and with previously performed calculations.
\end{abstract}
%

\maketitle

\tableofcontents	
\section{\label{Introduction}Introduction} 
It is well-known now that late transition-metal nitrides (TMNs) usually possess interesting properties leading to a variety of potential technological applications \cite{PhysRevB.82.104116,OsCarbides_2010,MorenoArmenta2007166}. Hence, a significant number of quantum mechanical \textit{ab initio} calculations of the structural and physical properties of this family of materials have appeared in the literature.

Since Juza and Hahn \cite{the_first_Cu3N_1939_exp} succeeded to synthesize Cu$_3$N in 1939, copper nitrides have been produced through various techniques and their properties and applications have been the subject of many theoretical and experimental published works \cite{Suleiman_PhD_arXiv2012_copper_nitrides_article}. Due to its early discovery, copper nitride may now be considered as the most investigated among the late TMNs \cite{nano_Cu3N_2005_exp}.

On the other hand, the nitride of silver, the next element to copper in group 11 of the periodic table, has been known for more than two centuries \cite{Ag3N_1991_exp,Ag_Pd_Au_nitrides_PhD_thesis_2010_exp}. However, despite its earlier discovery, silver nitride may be the least theoretically studied solid in the late TMNs family. Experimental efforts to investigate
 structural \cite{German_Ag3N_structure_1949_exp,Ag_Pd_Au_nitrides_PhD_thesis_2010_exp},
 electronic \cite{Ag_Pd_Au_nitrides_PhD_thesis_2010_exp} and
 formation \cite{Ag3N_1991_exp,Anderson_n_Parlee_1970_exp,German_Ag3N_structure_1949_exp,Juza_n_Hahn_1940}
 properties of silver nitrides have been made by some researchers.

In 1949, Hahn and Gilbert \cite{German_Ag3N_structure_1949_exp} carried out the first \cite{Ag_Pd_Au_nitrides_PhD_thesis_2010_exp} structural study on the reported stoichiometry, Ag$_3$N. They claimed an fcc structure with $a = 4.369 \; \text{\AA}$ and $Z = 4/3$ (i.e. $4$ Ag atoms in the unit cell). A long time later in 1982, Haisa \cite{Ag3N_structure_theory_1982} suggested that the Ag atoms are located at the corners and face centers of the unit cell, while the N atoms, which may be statistically distributed in the octahedral interstices, were given no definite positions \cite{Ag3N_structure_theory_1982}.

According to the calculated N radius, Ag$_3$N can be described as an ionic compound, and recent \textit{ab initio} calculations on the proposed structure revealed insulating characteristics with a fundamental band gap close to $1.35 \; eV$. On the other hand, due to the similar lattice of the parent Ag and the easily separated N as N$_2$, it can also be argued that this compound is a metal, supporting its black color \cite{Ag_Pd_Au_nitrides_PhD_thesis_2010_exp}.

Under ordinary conditions \cite{Ag3N_1991_exp}, it was found that silver can form Ag$_3$N
 \footnote{Ag$_3$N, formerly termed \textit{fulminating silver} by its discoverers, can be formed from ammoniacal solutions of silver oxide according to the following reaction
\begin{eqnarray} \label{fulminating silver reaction} 3 \text{Ag}_{2}\text{O} + 2 \text{N}\text{H}_3^{\text{(aq)}} \longrightarrow  2 \text{Ag}_3\text{N} + 5 \text{H}_2\text{O}.
\end{eqnarray} It can also be formed by means of other reactions \cite{Ag3N_1991_exp,Ag_Pd_Au_nitrides_PhD_thesis_2010_exp}.}
 from ammoniacal solutions of silver oxide \cite{Ag3N_1991_exp,Ag_Pd_Au_nitrides_PhD_thesis_2010_exp}.
The black metallic-looking solid outcome, Ag$_3$N, is an extremely sensitive explosive compound \cite{Ag3N_1991_exp,Hazardous_Silver_Compounds_1991}. It may explode due to the slightest touch, even from the impact of a falling water droplet \cite{Hazardous_Silver_Compounds_1991}, but it is relatively easy to handle under water or ethanol \cite{Ag_Pd_Au_nitrides_PhD_thesis_2010_exp}. The explosive power is due to the energy released during the decomposition reaction:
\begin{eqnarray} \label{Ag3N decomposition reaction}
2 \text{Ag}_3\text{N} \longrightarrow 6 \text{Ag} + \text{N}_2 \; .
\end{eqnarray}
Even in storage at room temperature, this solid compound decomposes slowly according to Eq. \ref{Ag3N decomposition reaction} above \cite{Hazardous_Silver_Compounds_1991,Ag_Pd_Au_nitrides_PhD_thesis_2010_exp}. From a  thermochemical point of view, it was found that there is no stable intermediate stage in this decomposition, but there may be a metastable intermediate species (phase) with a remarkably low decomposition rate \cite{Ag3N_1991_exp}. At this point, it may be worth mentioning that the thermochemistry of silver nitride systems is not fully documented in standard handbook data \cite{Ag3N_1991_exp}.

In their 1991 work, Shanley and Ennis \cite{Ag3N_1991_exp} stated: \textit{``Many of the samples ... did not survive the minimum handling required to move them, container and all, to the X-ray stage. ... More vigorously explosive samples propagated throughout their mass leaving no visible residue. Even among supersensitive materials, silver nitride is a striking example of a compound ``teetering on the edge of existence". Under the circumstances, we did not succeed in developing data on the proportion of silver nitride required for explosive behavior in these mixtures."}

Thus, beside the potential hazard to lab workers due to its sensitive explosive behavior, characterization of silver nitride is hindered by its extremely unstable (endothermic) nature \cite{Ag3N_1991_exp,Ag_Pd_Au_nitrides_PhD_thesis_2010_exp}, and we are presented with an incomplete picture of structural, electronic and optical properties of this material. Surprisingly, this lack of detailed knowledge of many physical properties of silver nitride stimulated only very few published \textit{ab initio} studies. 

In the present work, first-principles calculations were carried out to investigate the lattice parameters, equation of state, relative stabilities, phase transitions, electronic and optical properties of silver nitrides in three different chemical formulae and in various crystal structures. Calculation methods are described in Sec. \ref{Calculation Methods}. In Sec. \ref{Results and Discussion}, results are presented, discussed and compared with experiment and with previous calculations. The article is concluded with some remarks in Sec. \ref{Conclusions}.
%
\section{Calculation Methods}\label{Calculation Methods}
%
\subsection{\label{Stoichiometries and Crystal Structures}Stoichiometries and Crystal Structures}
To the best of our knowledge, the only experimentally reported stoichiometries of Ag-N compounds are Ag$_3$N \cite{Ag3N_1991_exp} and AgN$_3$ \cite{Ag3N_1991_exp}. However, previous \textit{ab initio} studies on Ag-N compounds considered
 Ag$_4$N \cite{Ag_Pd_Au_nitrides_PhD_thesis_2010_exp},
 Ag$_3$N \cite{AgN_AgN2_Ag2N_Ag3N_2010_comp,Ag_Pd_Au_nitrides_PhD_thesis_2010_exp},
 Ag$_2$N \cite{AgN_AgN2_Ag2N_Ag3N_2010_comp},
 AgN \cite{AgN_AgN2_Ag2N_Ag3N_2010_comp,CuN_AgN_AuN_2007_comp} and
 AgN$_2$ \cite{AgN2_AuN2_PtN2_2005_comp,AgN_AgN2_Ag2N_Ag3N_2010_comp}
in some cubic structures only. Consideration of stoichiometries other than the reported ones is probably due to the fact that many transition metals nitrides (TMs) are known to form more than one nitride \cite{StructuralInChem}. Hence, our interest in investigating AgN and AgN$_2$ is based on this fact.

For Ag$_3$N, we consider the following seven structures:
the face-centered cubic structure of AlFe$_3$ (D0$_3$),
the simple cubic structure of Cr$_3$Si (A15),
the simple cubic structure of the anti-ReO$_3$ (D0$_9$),                    
the simple cubic structure of Ag$_3$Au (L1$_2$),               
the body-centered cubic structure of CoAs$_3$ (D0$_2$),
the hexagonal structure of $\epsilon$-Fe$_3$N,
and the trigonal (rhombohedric) structure of RhF$_3$.    

For AgN, the following four structures were considered:
the face-centered cubic structure of NaCl (B1),
the simple cubic structure of CsCl (B2),
the face-centered cubic structure of ZnS zincblende (B3),
the hexagonal structure of NiAs (B8$_{1}$),
the hexagonal structure of BN (B$_{\text{k}}$),
the hexagonal structure of WC (B$_{\text{h}}$),
the hexagonal structure of ZnS wurtzite (B4),
the simple tetragonal structure of PtS cooperite (B17),
and the face-centered orthorhombic structure of TlF (B24).  

AgN$_2$ was studied in the following nine structures:
the face-centered cubic structure of CaF$_{2}$ fluorite (C1),
the simple cubic structure of FeS$_{2}$ pyrite (C2),    
the simple orthorhombic structure of FeS$_{2}$ marcasite (C18)
and the simple monoclinc structure of CoSb$_{2}$ (CoSb$_{2}$).
%
\subsection{\label{Electronic Relaxation Details}Electronic Relaxation Details}
In this work, electronic structure spin density functional theory (SDFT) \cite{SDFT_1972,SDFT_Pant_1972} calculations as implemented in the VASP code\cite{Vasp_ref_PhysRevB.47.558_1993,Vasp_ref_PhysRevB.49.14251_1994,Vasp_cite_Kressw_1996,
Vasp_PWs_Kresse_1996,DFT_VASP_Hafner_2008,PAW_Kresse_n_Joubert} have been employed. To self-consistently solve the Kohn-Sham (KS) equations \cite{KS_1965}
\begin{eqnarray}	\label{KS equations}
\begin{split}
  \Bigg \{ - \frac{\hbar^{2}} {2m_{e}}  \nabla^{2} + \int d\mathbf{r}^{\prime} \frac{n(\mathbf{r}^{\prime})}{|\mathbf{r}-\mathbf{r}^{\prime}|} + V_{ext}(\mathbf{r}) \\  + V_{XC}^{\sigma, \mathbf{k}}[n(\mathbf{r})] \Bigg \} \psi_{i}^{\sigma, \mathbf{k}}(\mathbf{r})  =  
   \epsilon_{i}^{\sigma, \mathbf{k}} \psi_{i}^{\sigma, \mathbf{k}}(\mathbf{r}),
\end{split}
\end{eqnarray}
where $i$, $\mathbf{k}$ and $\sigma$ are the band, $\mathbf{k}$-point and spin indices, receptively, VASP expands the pseudo part of the KS one-particle spin orbitals $\psi_{i}^{\sigma , \mathbf{k}}(\mathbf{r})$ on a basis set of plane-waves (PWs). Only those PWs with cut-off energy $E_{cut} \leq 600 \; eV$ have been included. The Brillouin zones were sampled using $\mathbf{\Gamma}$-centered Monkhorst-Pack \cite{MP_k_mesh_1976} $17 \times 17 \times 17$ meshes. Any increase in the $E_{cut}$ value or in the density of the $\mathbf{k}$-mesh produces a change in the total energy less than $3 \; \text{m} eV/ \text{atom}$. For static calculations, partial occupancies were set using the tetrahedron method with Bl\"{o}chl corrections \cite{tetrahedron_method_theory_1971,tetrahedron_method_theory_1972,ISMEAR5_1994}, while the smearing method of Methfessel-Paxton (MP) \cite{MP_smearing_1989} was used in the ionic relaxation, and Fermi surface of the metallic phases has been carefully treated. The Perdew-Burke-Ernzerhof (PBE) parametrization \cite{PBE_GGA_1996,PBE_GGA_Erratum_1997,XC_PBE_1999} of the generalized gradient approximation (GGA) \cite{XC_GGA_1988,XC_GGA_applications_1992,XC_GGA_applications_1992_ERRATUM} was employed
 for the exchange-correlation potentials $V_{XC}^{\sigma, \mathbf{k}}[n(\mathbf{r})]$. The implemented projector augmented wave (PAW) method\cite{PAW_Blochl, PAW_Kresse_n_Joubert} was used to describe the core-valence interactions $V_{ext}(\mathbf{r})$, where the $4d^{10}5s^{1}$ electrons of Ag and the $2s^{2}2p^{3}$  electrons of N are treated explicitly as valence electrons. While for these valence electrons only scalar kinematic relativistic effects are incorporated, the PAW potential treats the core electrons in a fully relativistic fashion\cite{DFT_VASP_Hafner_2008}. No spin-orbit interaction of the valence electrons has been considered.
%
\subsection{\label{Geometry Relaxation and EOS}Geometry Relaxation and EOS} 
At a set of externally imposed lattice constants, ions with free internal parameters were relaxed until all force components on each ion were less than $0.01 \; eV/\text{\AA}$. This is done following the implemented conjugate-gradient (CG) algorithm. After each ion relaxation step, static total energy calculation with the tetrahedron method was performed, and the cohesive energy per atom $E_{coh}$ was calculated from
\begin{eqnarray} \label{E_coh equation}
E_{coh}^{\text{Ag}_{m}\text{N}_{n}}  =   \frac{  E_{\text{solid}}^{\text{Ag}_{m}\text{N}_{n}} - Z \times \left( m E_{\text{atom}}^{\text{Ag}} + n E_{\text{atom}}^{\text{N}} \right) }{Z \times (m + n)},
\end{eqnarray}
where $Z$ is the number of Ag$_{m}$N$_{n}$ formulas per unit cell, $E_{\text{atom}}^{\text{Ag}}$ and $E_{\text{atom}}^{\text{N}}$ are the energies of the isolated non-spherical spin-polarized atoms, and $m,n = 1,2 \text{ or } 3$ are the stoichiometric weights. The obtained $E_{coh}$ as a function of volume $V$ per atom were then fitted to a Birch-Murnaghan 3rd-order equation of state (EOS)\cite{BM_3rd_eos} and the equilibrium volume $V_{0}$, the equilibrium cohesive energy $E_{0}$, the equilibrium bulk modulus
\begin{equation}	\label{B_0 eq}
B_{0} = -V \frac{\partial P}{\partial V}\Big|_{V=V_{0}} = -V \frac{\partial^{2} E}{\partial V^{2}}\Big|_{V=V_{0}}
\end{equation}
and its pressure derivative
\begin{equation}	\label{B^prime eq}
 B^{\prime}_{0} = \frac{\partial B}{\partial P} \Big|_{P=0} = \frac{1}{B_{0}} \left(  V \frac{\partial}{\partial V} (V \frac{\partial^{2} E}{\partial V^{2}}) \right)  \Big|_{V=V_{0}} 	
\end{equation}
were determined.
%
\subsection{Formation Energy}\label{Formation Energy}
Beside the cohesive energy, another measure of relative stability is the formation energy $E_f$. Assuming that silver nitrides Ag$_m$N$_n$ result from the interaction between the N$_{2}$ gas and the solid Ag(A1) via the reaction (compare with Eq. \ref{Ag3N decomposition reaction})
\begin{eqnarray} \label{E_f reaction}
m \text{Ag}^{\text{solid}} + \frac{n}{2} \text{N}_2^{\text{gas}} \rightleftharpoons   \text{Ag}_m\text{N}_n^{\text{solid}},
\end{eqnarray}
$E_f$ can be given by
\begin{align} \label{formation energy equation}
E_f(\text{Ag}_m\text{N}_n^{\text{solid}}) =   E_\text{coh}(\text{Ag}_m\text{N}_n^{\text{solid}}) \quad \quad \quad \quad \quad \quad &
\nonumber \\
- \frac{  m E_\text{coh}(\text{Ag}^{\text{solid}}) + \frac{n}{2} E_\text{coh}(\text{N}_2^{\text{gas}})}{m + n} &		\; .
\end{align}
Here $m,n=1,2,3$ are the stoichiometric weights and $E_\text{coh}(\text{Ag}_m\text{N}_n^{\text{solid}})$ is the cohesive energy per atom as in Eq. \ref{E_coh equation}. The cohesive energy $E_\text{coh}(\text{Ag}^{\text{solid}})$ and other equilibrium properties of the elemental metallic silver are given in Table \ref{silver_nitrides_equilibrium_structural_properties}. The cohesive energy of the diatomic nitrogen ($E_\text{coh}(\text{N}_2^{\text{gas}})$) was found to be $-5.196 \; eV/\text{atom}$ corresponding to an equilibrium  N--N bond length of $1.113 \; \text{\AA}$ (For more details, see Ref. \onlinecite{Suleiman_PhD_arXiv2012_copper_nitrides_article}).
%
\subsection{\label{GWA Calculations and Optical Properties}GWA Calculations and Optical Properties}
Although a qualitative agreement between DFT-calculated optical properties and experiment is possible, accurate quantitative description requires treatments beyond the level of DFT \cite{PAW_optics}. Another approach provided by many-body perturbation theory (MBPT) leads to a system of quasi-particle (QP) equations, which can be written for a periodic crystal as \cite{GWA_and_QP_review_1999,Kohanoff,JudithThesis2008}
\begin{eqnarray}	\label{QP equations}
\begin{split}
  \Bigg \{ - \frac{\hbar^{2}} {2m}  \nabla^{2} + \int d\mathbf{r}^{\prime} \frac{n(\mathbf{r}^{\prime})}{|\mathbf{r}-\mathbf{r}^{\prime}|} + V_{ext}(\mathbf{r}) \Bigg \} \psi_{i,\mathbf{k}}^{QP}(\mathbf{r}) \\  + \int d\mathbf{r}^{\prime} \Sigma(\mathbf{r},\mathbf{r}^{\prime};\epsilon_{i,\mathbf{k}}^{QP})  \psi_{i,\mathbf{k}}^{QP}(\mathbf{r}^{\prime}) = \epsilon_{i,\mathbf{k}}^{QP} \psi_{i,\mathbf{k}}^{QP}(\mathbf{r}).
\end{split}
\end{eqnarray}
Practically, the wave functions $\psi_{i,\mathbf{k}}^{QP}(\mathbf{r})$ are taken from the DFT calculations. However, in consideration of computational cost, we used a less dense mesh of $\mathbf{k}$-points ($10 \times 10 \times 10$). The term $\Sigma(\mathbf{r},\mathbf{r}^{\prime};\epsilon_{i,\mathbf{k}}^{QP})$ is the self-energy which contains all the exchange and correlation effects, static and dynamic, including those neglected in our DFT-GGA reference system. In the so-called $GW$ approximation \cite{Hedin_1st_GWA_1965}, $\Sigma$ is given in terms of the Green's function $G$ as
\begin{eqnarray}	\label{GW self-energy}
\begin{split}
\Sigma_{GW} = j \int d\epsilon^{\prime} G(\mathbf{r},\mathbf{r}^{\prime};\epsilon,\epsilon^{\prime}) W(\mathbf{r},\mathbf{r}^{\prime};\epsilon),
\end{split}
\end{eqnarray}
where the dynamically (frequency dependent) screened Coulomb interaction $W$ is related to the bare Coulomb interaction $v$ through 
\begin{eqnarray}
\begin{split}
W(\mathbf{r},\mathbf{r}^{\prime};\epsilon) = j \int d\mathbf{r}_{1} \varepsilon^{-1}(\mathbf{r},\mathbf{r}_{1};\epsilon)v(\mathbf{r}_{1},\mathbf{r}^{\prime}),
\end{split}
\end{eqnarray}
with $\varepsilon$, the dielectric matrix, calculated within the so-called random phase approximation (RPA). We followed the $GW_{0}$ self-consistent routine on $G$, in which the QP eigenvalues
\begin{eqnarray} \label{QP eigenvalues}
\begin{split}
\epsilon_{i,\mathbf{k}}^{QP} = \text{Re}  \left( 
\left\langle \psi_{i,\mathbf{k}}^{QP} \middle| 
H_{\text{KS}} - V_{XC} + \Sigma_{GW_{0}}
\middle|  \psi_{i,\mathbf{k}}^{QP} \right\rangle  	\right)
\end{split}
\end{eqnarray}
are updated in the calculations of $G$, while $W$ is kept at the DFT-RPA level. Four updates were performed, and after the final iteration in $G$, $\varepsilon$ is recalculated within the RPA using the updated QP eigenvalues \cite{Kohanoff,JudithThesis2008,VASPguide}. From the real $\varepsilon_{\text{re}}(\omega)$ and the imaginary $\varepsilon_{\text{im}}(\omega)$ parts of this frequency-dependent microscopic dielectric tensor, one can derive all the other frequency-dependent dielectric response functions, such as
 reflectivity $R\left(\omega\right)$,
 transmitivity $T\left(\omega\right) = 1- R\left(\omega\right)$,
 refractive index $n\left(\omega\right)$,
 extinction coefficient $\kappa\left(\omega\right)$, and
 absorption coefficient $\alpha\left(\omega\right)$ \cite{Fox,dressel2002electrodynamics,Ch9_in_Handbook_of_Optics_2010}: 
\begin{align}
R\left(\omega\right) &= \left| \frac{\left[  \varepsilon_{\text{re}}\left(\omega\right) + j \varepsilon_{\text{im}}\left(\omega\right)  \right]^{\frac{1}{2}} - 1}{\left[  \varepsilon_{\text{re}}\left(\omega\right) + j \varepsilon_{\text{im}}\left(\omega\right)  \right]^{\frac{1}{2}} + 1} \right| ^{2}		\label{R(omega)}\\
n\left(\omega\right) &= \frac{1}{\sqrt{2}} \left(  \left[  \varepsilon_{\text{re}}^{2}\left(\omega\right) + \varepsilon_{\text{im}}^{2}\left(\omega\right) \right]^{\frac{1}{2}} + \varepsilon_{\text{re}}\left(\omega\right) \right)^{\frac{1}{2}} 			\label{n(omega)}\\
\kappa\left(\omega\right) &= \frac{1}{\sqrt{2}} \left(  \left[  \varepsilon_{\text{re}}^{2}\left(\omega\right) + \varepsilon_{\text{im}}^{2}\left(\omega\right) \right]^{\frac{1}{2}} - \varepsilon_{\text{re}}\left(\omega\right) \right)^{\frac{1}{2}} 			\label{kappa(omega)}\\
\alpha\left(\omega\right) &= \sqrt{2} \omega  \left(  \left[  \varepsilon_{\text{re}}^{2}\left(\omega\right) + \varepsilon_{\text{im}}^{2}\left(\omega\right) \right]^{\frac{1}{2}}  - \varepsilon_{\text{re}}\left(\omega\right) \right)^{\frac{1}{2}} 		\label{alpha(omega)}
\end{align}
%
\section{Results and Discussion}\label{Results and Discussion}
The energy-volume equation of state (EOS) for the different structures of Ag$_3$N, AgN$_2$ and AgN are depicted in Figs. \ref{Ag3N1_ev_EOS}, \ref{Ag1N1_ev_EOS} and \ref{Ag1N2_ev_EOS}, respectively. The corresponding calculated  equilibrium properties are given in Table \ref{silver_nitrides_equilibrium_structural_properties}. In this table, we ordered the studied phases according to the increase in the nitrogen content; then within each series, structures are ordered in the direction of decreasing structural symmetry. For the sake of comparison, we also presented results from experiment and from previous \textit{ab initio} calculations; and, whenever appropriate, the calculation method and the $XC$ functional are also given in footnotes of the Table.

The calculated equilibrium properties: cohesive energies, formation energies, volume per atom, volume per Ag atom, and bulk modulus and its pressure derivative which are given Table \ref{silver_nitrides_equilibrium_structural_properties}, are visualized in Fig. \ref{silver_nitrides_equilibrium_properties}. This kind of visualization allows us to study the effect of nitridation on the parent Ag(A1), since all quantities in this figure are given relative to the corresponding ones of the elemental Ag(A1) given in the first row of Table \ref{silver_nitrides_equilibrium_structural_properties}. Moreover, one can easly compare the properties of these phases relative to each other.\\
%
\subsection{EOS and Relative Stabilities}\label{EOS and Relative Stabilities}
%
\begin{figure}[!]
\includegraphics[width=0.45\textwidth]{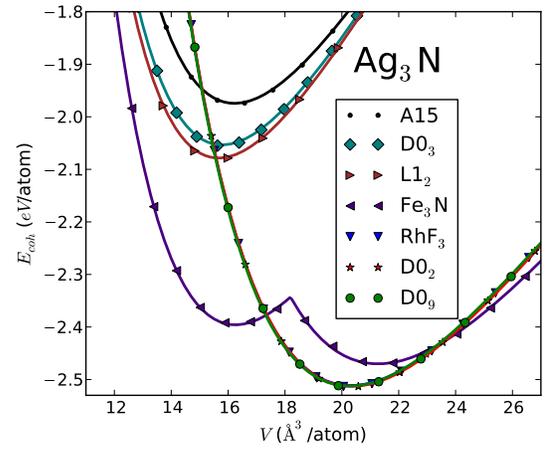}
\caption{\label{Ag3N1_ev_EOS}(Color online.) Cohesive energy $E_{\text{coh}} (eV/\text{atom})$ versus atomic volume $V$ (\AA$^{3}$/\text{atom}) for Ag$_3$N in seven different structural phases.}
\end{figure}
\begin{figure}[!]
\includegraphics[width=0.45\textwidth]{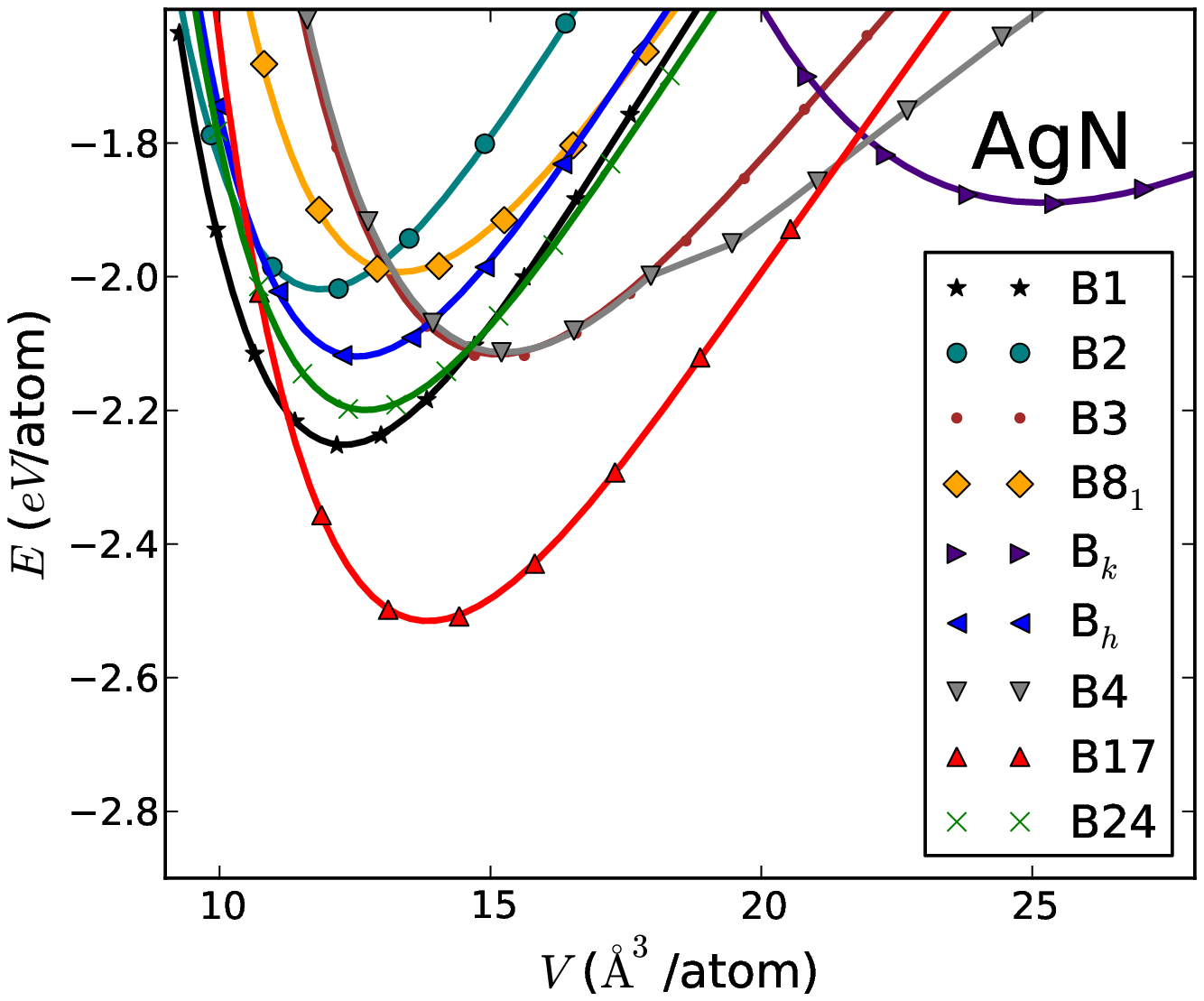}
\caption{\label{Ag1N1_ev_EOS}(Color online.) Cohesive energy $E_{\text{coh}} (eV/\text{atom})$ versus atomic volume $V$ (\AA$^{3}$/\text{atom}) for AgN in nine different structural phases.}
\end{figure}
\begin{figure}[!]
\includegraphics[width=0.45\textwidth]{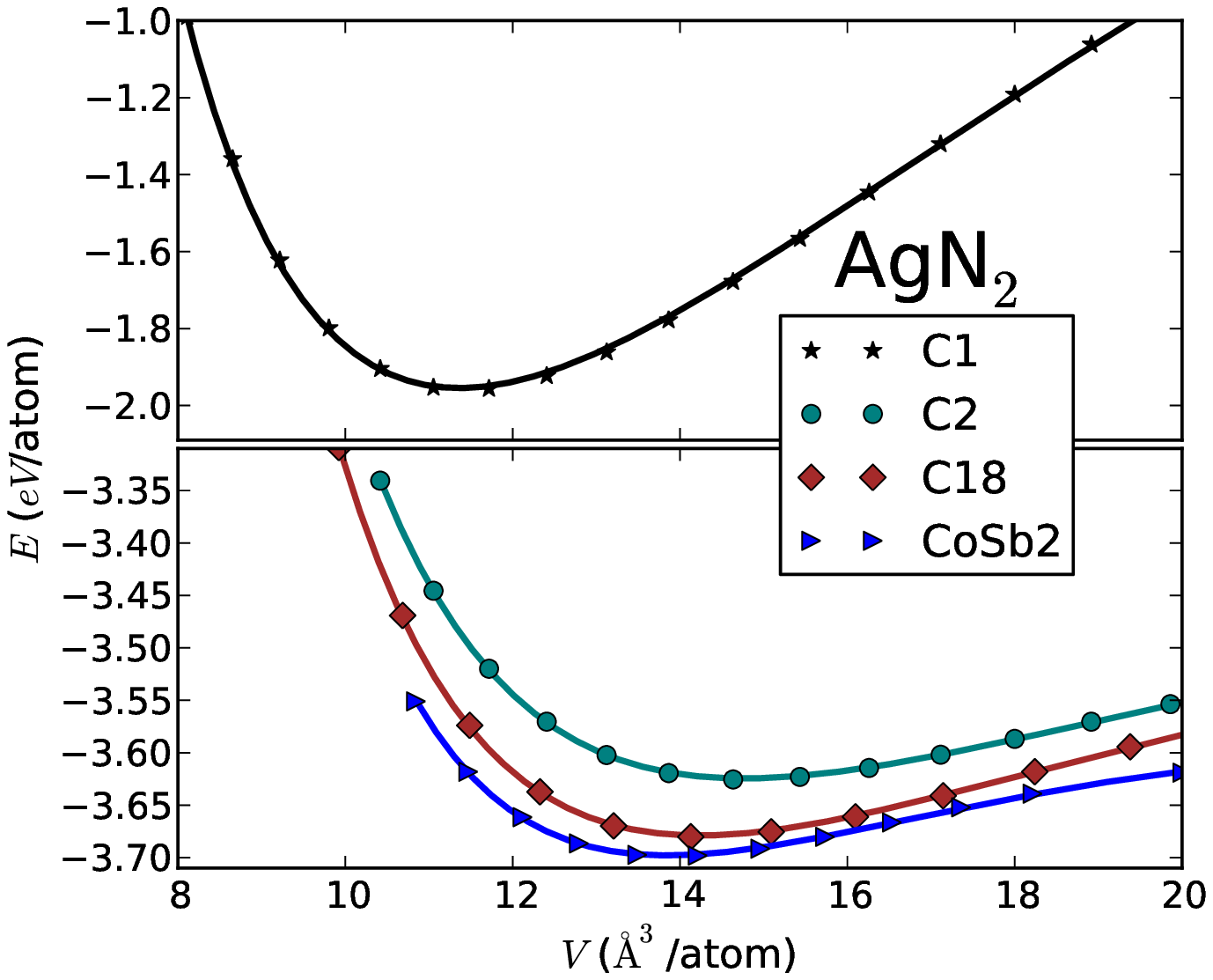}
\caption{\label{Ag1N2_ev_EOS}(Color online.) Cohesive energy $E_{coh} (eV/\text{atom})$ versus atomic volume $V (\AA^{3}/\text{atom})$ for AgN$_2$ in four different structural phases.}
\end{figure}
%
\input{footed_results_table.tex}
%
\begin{figure*}[H!]
\includegraphics[width=0.95\textwidth]{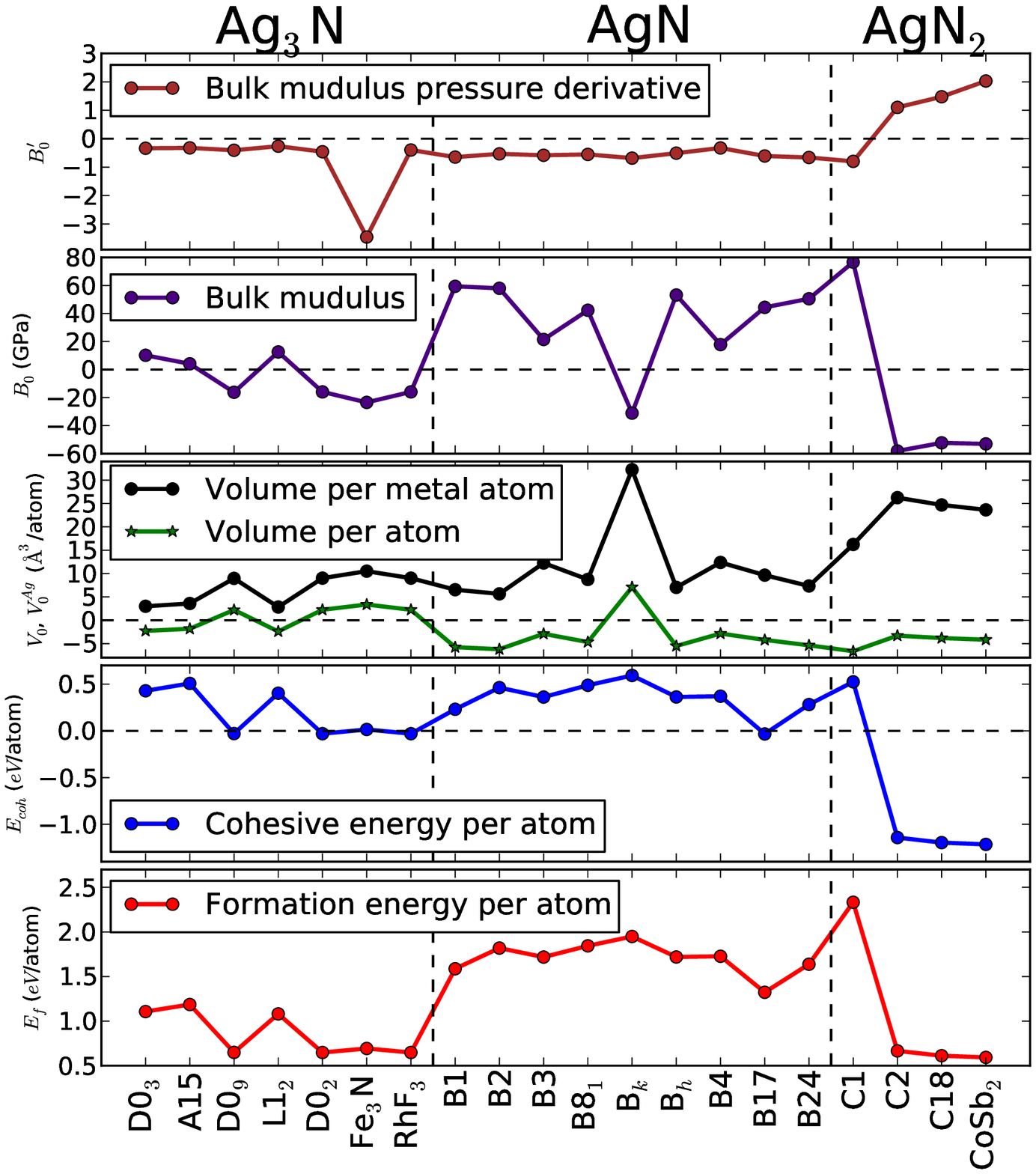}
\caption{\label{silver_nitrides_equilibrium_properties}(Color online.) Calculated equilibrium properties of the twenty studied phases of silver nitrides. All quantities are given relative to the corresponding ones of the \textit{fcc} crystalline elemental silver given in the first row of Table \ref{silver_nitrides_equilibrium_structural_properties}. The vertical dashed lines separate between the different stoichoimetries.}
\end{figure*}
%
Considering $E_{\text{coh}}$ in the Ag$_3$N series, Fig. \ref{Ag3N1_ev_EOS} shows clearly that the $E(V)$ relations of Ag$_3$N in D0$_9$, D0$_2$ and RhF$_3$ phases are almost identical, corresponding to equilibrium cohesive energy (Table \ref{silver_nitrides_equilibrium_structural_properties}) of $-2.513$, $-2.514$ and $-2.514 \;\; eV/\text{atom}$, respectively. This behavior in the EOS could be traced back to the structural relationships between these three structures, since both D0$_{2}$ and RhF$_{3}$ can simply be derived from the more symmetric D0$_{9}$ (see Ref. \onlinecite{Suleiman_PhD_arXiv2012_copper_nitrides_article} and references therein). These structural relations may reflect in the EOS's and in other physical properties, and the three phases may co-exist during the Ag$_3$N synthesis process. Relative to the elemental Ag, these three phases tend not to change the $E_{\text{coh}}$ (Fig. \ref{silver_nitrides_equilibrium_properties}), lowering it only by $\sim 0.03 \;eV/\text{atom}$, as can been seen from Table \ref{silver_nitrides_equilibrium_structural_properties}. It may be worth to mention here that the simple cubic D0$_{9}$ phase is the stable phase of the synthesized Cu$_3$N \cite{the_first_Cu3N_1939_exp,Cu3N_1996_comp}.

The odd behavior of the EOS of Ag$_3$N(Fe$_3$N) with the existence of two minima (Fig. \ref{Ag3N1_ev_EOS}) shows that the first minima (to the left) is a metastable local minimum that cannot be maintained as the system is decompressed. Ag ions are in the $6g$ Wyckoff positions: $(x,0,0), (0,x,0), (-x,-x,0), (-x,0,\frac{1}{2}), (0,-x,\frac{1}{2})$ and $(x,x,\frac{1}{2})$; with $x \sim \frac{1}{3}$ to the left of the potential barrier (represented by the sharp peak at $\sim 18.2 \; \text{\AA}^3/\text{atom}$), and $x=\frac{1}{2}$ to the right of the peak.  It may be relevant to mention here that Cu$_3$N(Fe$_3$N) was found to behave in a similar manner \cite{Suleiman_PhD_arXiv2012_copper_nitrides_article}.

\begin{figure}[!]
\includegraphics[width=0.45\textwidth]{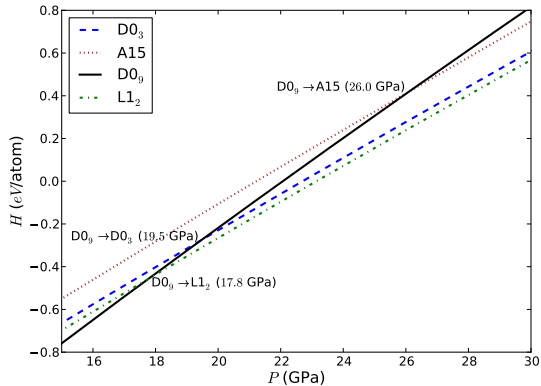}
\caption{\label{Ag3N1_hp_EOS}(Color online.) Enthalpy $H$ \textit{vs.} pressure $P$ equation of state (EOS) for some Ag$_3$N phases in the range where D0$_9$ $\rightarrow$ A15, D0$_9$ $\rightarrow$ D0$_3$ and D0$_9$ $\rightarrow$ L1$_2$ phase transitions occur.}
\end{figure}

The crossings of the less stable A15, D0$_3$ and L1$_2$ EOS curves with the more stable D0$_9$, D0$_2$ and RhF$_3$ ones at the left side of their  equilibrium points reveals pressure-induces phase transitions from the latter phases to the former. To show this, we plotted the corresponding relation between enthalpy $H = E(V) + PV$ and the imposed pressure $P$ in Fig. \ref{Ag3N1_hp_EOS}. Since D0$_9$, D0$_2$ and RhF$_3$ phases have identical $E(V)$ curves, the corresponding $H(P)$ curves are also identical. Hence, only the $H(P)$ of D0$_9$ is displayed in Fig. \ref{Ag3N1_hp_EOS}. A point where the enthalpies of two phases are equal determine the phase transition pressure $P_{t}$; and, indeed, the direction of the transition is from the higher $H$ to the lower $H$ \cite{Grimvall}. As depicted in Fig. \ref{Ag3N1_hp_EOS}, $P_{t}(\text{D0}_9 \rightarrow \text{L1}_2) = 17.8 \; \text{GPa}$, $P_{t}(\text{D0}_9 \rightarrow \text{D0}_3) = 19.5 \; \text{GPa}$ and $P_{t}(\text{D0}_9 \rightarrow \text{A15}) = 26.0 \; \text{GPa}$. Thus, D0$_9$, D0$_2$ and RhF$_3$ would not survive behind these $P_{t}$'s and A15, D0$_3$ and L1$_2$ are preferred at high pressure.

%
%
Fig. \ref{silver_nitrides_equilibrium_properties} reveals that the AgN group contains the least stable phase among all the twenty studied phases: the hexagonal B$_{\text{k}}$. Fig. \ref{Ag1N1_ev_EOS} and Table \ref{silver_nitrides_equilibrium_structural_properties} show that the simple tetragonal structure of cooperite (B17) is the most stable phase in this AgN series. In fact, one can see from Fig. \ref{silver_nitrides_equilibrium_properties} and Table \ref{silver_nitrides_equilibrium_structural_properties} that all the considered AgN phases possess less binding than their parent Ag(fcc), except AgN(B17) which is slightly more stable, with $0.033 \; eV/\text{atom}$ lower $E_{\text{coh}}$. It is interesting to notice that AgN(B17) is $\sim 0.003 \; eV/\text{atom}$ more stable than the Ag$_3$N most stable phases. Moreover, this B17 structure was theoretically predicted to be the ground-state structure of CuN \cite{Suleiman_PhD_arXiv2012_copper_nitrides_article}, AuN \cite{Suleiman_PhD_SAIP2012_gold_nitrides_article}, PdN \cite{Suleiman_PhD_SAIP2012_palladium_nitrides_article} and PtN \cite{PtN_2006_comp_B17_structure_important}.

Using the full-potential (linearized) augmented plane waves plus local orbitals (FP-LAPW+lo) method within LDA and within GGA, Kanoun and Said \cite{CuN_AgN_AuN_2007_comp} studied the $E(V)$ EOS for AgN in the B1, B2, B3 and B4 structures. The equilibrium energies they obtained from the $E(V)$ EOS revealed that B1 is the most stable phase, and the relative stability they arrived at is in the order B1--B3--B4--B2, with a significant difference in total energy between B3 and B4 (see Fig. 2(b) in that article). Within this subset of structures, the numerical values of $E_{\text{coh}}$ in Table \ref{silver_nitrides_equilibrium_structural_properties} do have the same order. However, the difference between the equilibrium $E_{\text{coh}}$(B3) and $E_{\text{coh}}$(B4) is only $\sim 0.009 \; eV$, and the $E(V)$ EOS of B3 and B4 match/overlap over a wide range of volumes around the equilibrium point. This discrepancy may be attributed to the unphysical/ill-defined measure of stability that Kanoun and Said used, the \textit{total} energy, while the number of the AgN formula units per unit cell in the B4 structure differs from that in the others\footnote{\textit{\textbf{In their original article\cite{CuN_AgN_AuN_2007_comp}, Kanoun and Said stated that ``... there are two \textit{atom} in wurtzite unit cell, and one in all the other cases." which is a clear typo!}}}. Nevertheless, it may be worth mentioning here that AgN(B3) was found to be elastically unstable \cite{AgN_AgN2_Ag2N_Ag3N_2010_comp}.

In the CuN$_2$ nitrogen-richest phase series, we can see from Table \ref{silver_nitrides_equilibrium_structural_properties} and from Fig. \ref{silver_nitrides_equilibrium_properties} that the phases of this group are significantly more stable than all the other studied phases, except C1, which is, in contrast, the second least stable among the twenty studied phases, with $0.017 \; eV/\text{atom}$ more than AgN(B$_{\text{k}}$). From Fig. \ref{Ag1N2_ev_EOS}, Fig. \ref{silver_nitrides_equilibrium_properties} and Table \ref{silver_nitrides_equilibrium_structural_properties}, one can see that in this series, the lower the structural symmetry, the more stable is the phase. It was found that CuN$_2$ phases have the same trend \cite{Suleiman_PhD_arXiv2012_copper_nitrides_article}.

Comparing the relative stability of Ag$_3$N, AgN and AgN$_2$, we find from Table \ref{silver_nitrides_equilibrium_structural_properties} and from Fig. \ref{silver_nitrides_equilibrium_properties} that AgN$_2$ in its least symmetric phase, the simple monoclinc structure of CoSb$_{2}$, is the most energetically stable phase with $E_{\text{coh}} = -3.699$ $eV/\text{atom}$.
%
\subsection{Volume per Atom and Lattice Parameters}\label{Volume per Atom and Lattice Parameters}
The numerical values of the lattice parameters and the average equilibrium volume per atom $V_{0}$ for the twenty modifications are presented in Table \ref{silver_nitrides_equilibrium_structural_properties}. The middle subwindow of Fig. \ref{silver_nitrides_equilibrium_properties} depicts the $V_{0}$ values relative to the Ag(fcc). To measure the average of  the Ag--Ag bond length in the silver nitride, the equilibrium average volume per Ag atom ($V_{0}^{Ag}$), which is simply the ratio of the volume the unit cell to the number of Ag atoms in the unit cell, is visualized in the same subwindow.

From the $V_{0}$ curve in Fig. \ref{silver_nitrides_equilibrium_properties}, we can see that, all AgN and AgN$_2$ modifications, except the open AgN(B$_{\text{k}}$) phase, decrease $V_{0}$; while the Ag$_3$N phases tend, on average, not to change the number density of the parent Ag(A1).

On the other hand, the $V_{0}^{\text{Ag}}$ curve in Fig. \ref{silver_nitrides_equilibrium_properties} reveals that, relative to the elemental Ag and to each other, $V_{0}^{Ag}$ \textit{tends} to increase with the increase in the nitrogen content. Thus, in all these nitrides, the introduced N ions displace apart the ions of the host lattice causing longer Ag-Ag bonds than in the elemental Ag. This cannot be seen directly from the $V_{0}$ values depicted in the same figure.

For AgN in the B1, B2, B3 and B4 structures, Kanoun and Said (Ref. \onlinecite{CuN_AgN_AuN_2007_comp} described in Sec. \ref{EOS and Relative Stabilities} above) obtained GGA equilibrium lattice parameters which are in very good agreement with ours. Their obtained LDA lattice parameter values show the common underestimation with respect to their and our GGA values (see Table \ref{silver_nitrides_equilibrium_structural_properties}).

Gordienko and Zhuravlev \cite{AgN_AgN2_Ag2N_Ag3N_2010_comp} studied the structural, mechanical and electronic properties of AgN(B1) , AgN(B2), AgN(B3), AgN$_2$(C1) and Ag$_3$N(D0$_9$) cubic phases. Their DFT calculations were based on pseudopotential (PP) method within LDA, and on linear combinations of atomic orbitals (LCAO) method within both LDA and GGA. For comparison, some of their findings are included in Table \ref{silver_nitrides_equilibrium_structural_properties}. Within the parameter subspace they considered, our GGA values of the $a$ lattice parameter agree very well with theirs. On the other hand, although their PP $a$ values are closer to the GGA ones (ours and theirs), all their LDA values are less than the GGA ones. This confirms the well-known behavior of LDA compared to GGA \cite{accurate_GGA_2006,GGA_vs_LDA_2004,Ch1_in_Primer_in_DFT_2003}. Gordienko and Zhuravlev also found that the Ag--Ag interatomic distance increases in the order Ag$_3$N--AgN--AgN$_2$. This agrees with the general trend shown in Fig. \ref{silver_nitrides_equilibrium_properties}, since the $V_{0}^{Ag}$ curve shows an average increase in the same direction.
%
\subsection{\label{Bulk Modulus and its Pressure Derivative}Bulk Modulus and its Pressure Derivative}
Fig. \ref{silver_nitrides_equilibrium_properties} reveals that Ag$_3$N phases tend, on average, to preserve the $B_{0}$ value of the parent Ag(A1). Increasing the nitrogen content to get AgN phases will increase the $B_{0}$ value of the parent Ag(A1), except in the case of B$_{k}$. While the nitrogen in AgN$_2$ tends to lower the $B_{0}$ value of the parent Ag(A1), the cubic C1 phase posses the highest $B_{0}$ value. This could be seen from Fig. \ref{Ag1N2_ev_EOS}, where the curvature of the $E_{\text{coh}}(V)$ curve of C1 is higher compared to the shallow minima of the C2, C18 and CoSb$_2$ curves.

From the definition of the equilibrium bulk modulus (Eq. \ref{B_0 eq}), one would expect $B_{0}$ to increase as $E_{\text{coh}}$ or $V_{0}$ decreases. This is because of the minus sign of the former and the inverse proportionality of the latter. That is, \textit{roughly speaking}, the $B_{0}$ curve should have a mirror reflection-like behavior with respect to the $E_{\text{coh}}$ and $V_{0}$ curves. Nevertheless, if $E_{\text{coh}}$ or $V_{0}$ are increasing and the other is decreasing, then the dominant net effect will be of the one with the higher change \footnote{\textbf{\textit{Since Eq. \ref{B_0 eq} does not refer to any stoichiometry or any species (that is, it does not consider the way that the change in energy or volume was done), we may take the change in volume (or energy) with respect to itself, with respect to the parent Ag(A1), or with respect to any of the other nineteen considered modifications.}}}. For example, Fig. \ref{silver_nitrides_equilibrium_properties} shows that in going from D0$_3$ to A15, both $E_{\text{coh}}$ and $V_{0}$ increase resulting in a negative change in $B_{0}$. In going from A15 to D0$_9$, $E_{\text{coh}}$ is decreasing while $V_{0}$ is increasing, but, in the end, the latter won the competition and lowered the value of $B_{0}$. This argument stays true throughout the three series. When there is no significant change in both $E_{\text{coh}}$ and $V_{0}$, there is no significant change in $B_{0}$. This is the case when one goes from C18 to CoSb$_{2}$. A close look at the $B_{0}$ curve in Fig. \ref{silver_nitrides_equilibrium_properties}, reveals that the huge decrease in $E_{\text{coh}}$ between C1 and C2 defeats the relatively small increase in $V_{0}$. This is simply because, according to Eq. \ref{B_0 eq}, the value of $B_{0}$ is proportional to the \textit{absolute} change in $E_{\text{coh}}$, while it is far more sensitive to any change in $V_{0}$ because it is proportional to $(\Delta V_{0})^{-1}$.

It is common to measure the pressure dependence of $B_{0}$ by its derivative $B^{\prime}_{0}$ (Eq. \ref{B^prime eq}). Fig. \ref{silver_nitrides_equilibrium_properties} shows that the $B_{0}$ value of the C2, C18 and CoSb$_2$ phases increases as these phases are put under pressure. While the $B_{0}$ values of the rest of the phases shows very low sensitivity to pressure and they tend to slightly lower the bulk modulus, the Fe$_3$N phase is the most sensitive phase and tends to significantly lower its $B_{0}$ upon application of pressure. This high sensitivity may indicate that the corresponding minimum on the potential surface is not global, but another local minimum as the one at $16.2 \; \AA^3/\text{atom}$ (Fig. \ref{Ag3N1_ev_EOS}).

From the elastic constants they obtained, Gordienko and Zhuravlev\cite{AgN_AgN2_Ag2N_Ag3N_2010_comp} calculated the corresponding macroscopic bulk moduli (included in Table \ref{silver_nitrides_equilibrium_structural_properties}). They found the highest LDA $B_{0}$ value for AgN(B1) among all phases they considered, but, in agreement with the present work, they obtained the highest GGA $B_{0}$ value for AgN$_2$(C1). Since LDA relative to GGA overestimates $E_{\text{coh}}$ and thus underestimates $V_{0}$, \textit{each} of these two factors (see Subsection \ref{Bulk Modulus and its Pressure Derivative}) would separately lead to the odd LDA value of $219.2$ GPa which they obtained. Nevertheless, due to this fact, Gordienko and Zhuravlev argued that one should consider the LDA and GGA average value of $B_{0}$.
%
\subsection{\label{Formation Energies}Formation Energies}
Formation energies in the present work are used as a measure of the \textit{relative} thermodynamic stabilities of the phases under consideration. That is, the lower the formation energy, the lower the tendency to dissociate back into the constituent components Ag and N$_2$.

The obtained formation energies $E_{f}$ of the twenty relaxed phases are given in Table \ref{silver_nitrides_equilibrium_structural_properties} and depicted graphically in Fig. \ref{silver_nitrides_equilibrium_properties}. The latter shows that, relative to each other and within each series, the formation energy $E_f$ (defined by Eqs. \ref{E_f reaction} and \ref{formation energy equation}) of the studied phases has the same trend as the cohesive energy\footnote{\textbf{\textit{Surely, this needs not to be so. Compare the definition \ref{E_coh equation} with the definition \ref{E_f reaction}.}}}. That is, all phases have the same relative stabilities in the $E_f$ space as in the $E_{\text{coh}}$ space. However, while Ag$_3$N phases tend to have equal $E_{\text{coh}}$ as the AgN phases, all Ag$_3$N modifications have a lower $E_f$ than the AgN ones. Hence, silver nitride is more likely to be formed in the former stoichiometry.
 However, all the twenty obtained $E_f$ values are positive; which, in principle, means that all these phases are thermodynamically unstable (endothermic) \footnote{\textbf{\textit{
It is common that one obtains positive DFT formation energy for (even the experimentally synthesized) transition-metal nitrides. Moreover, the zero-pressure zero-temperature DFT calculations have to be corrected for the conditions of formation of these nitrides. Another source of this apparent shortcoming stems from the PBE-GGA underestimation of the cohesion in N$_{2}$. We have discussed this point further in Ref. \onlinecite{Suleiman_PhD_arXiv2012_copper_nitrides_article}.}}}.

Some of the experimental values of $E_f$ for the synthesized Ag$_3$N phase (which is claimed to be in an fcc structure) are 
 $+199 \; \text{kJ/mol}$ \cite{Juza_n_Hahn_1940} = $2.062 \; eV/\text{atom}$,
 $+230 \; \text{kJ/mol}$ \cite{Anderson_n_Parlee_1970_exp} = $2.384 \; eV/\text{atom}$,
 $+255 \; \text{kJ/mol}$ \cite{German_Ag3N_structure_1949_exp} = $2.643 \; eV/\text{atom}$ and
 $(+314.4 \mp 2.5) \; \text{kJ/mol}$  \cite{Ag3N_1991_exp} = $(3.25853 \pm 0.02591) \; eV$;
 with an average value of $2.587 \pm 0.364 \; eV$. Among the considered phases in the present work, there is only one phase wich has $E_f$ value that fits in this range, the AgN$_2$(C1). Interestingly, this C1 structure has an fcc underlying Bravia lattice; however, the chemical formula differs from that of the synthesized phase.
%
\subsection{\label{Electronic Properties}Electronic Properties}
\begin{figure*}
\includegraphics[width=1.0\textwidth]{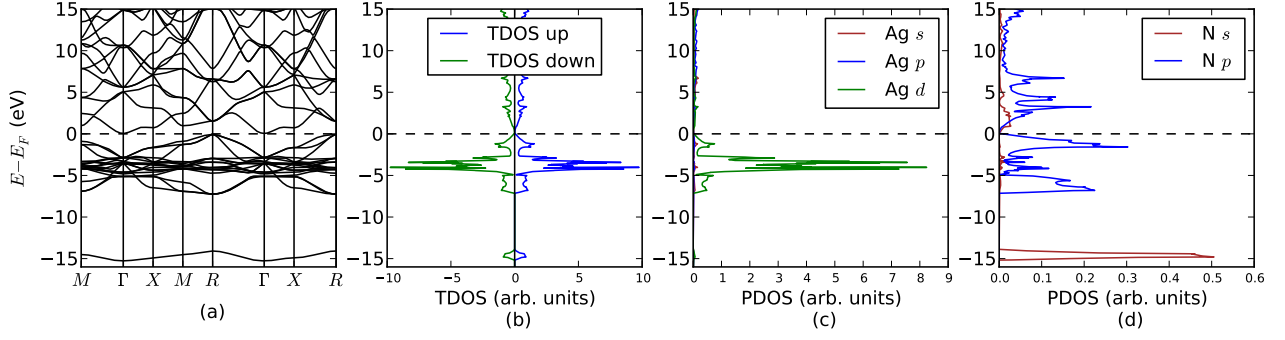}
\caption{\label{Ag3N1_D0_9_electronic_structure}(Color online.) DFT calculated electronic structure for Ag$_3$N in the D0$_9$ structure:
 \textbf{(a)} band structure along the high-symmetry $\mathbf{k}$-points which are labeled according to Ref. [\onlinecite{Bradley}]. Their coordinates w.r.t. the reciprocal lattice basis vectors are: $M(0.5, 0.5, 0.0)$, $\Gamma(0.0, 0.0, 0.0 )$, $X(0.0, 0.5, 0.0 )$, $R(0.5, 0.5, 0.5 )$;
 \textbf{(b)} spin-projected total density of states (TDOS);
 \textbf{(c)} partial density of states (PDOS) of Ag($s, p, d$) orbitals in Ag$_3$N;
 and \textbf{(d)} PDOS of N($s, p$) orbitals in Ag$_3$N,.}
\end{figure*}

\begin{figure*}
\includegraphics[width=1.0\textwidth]{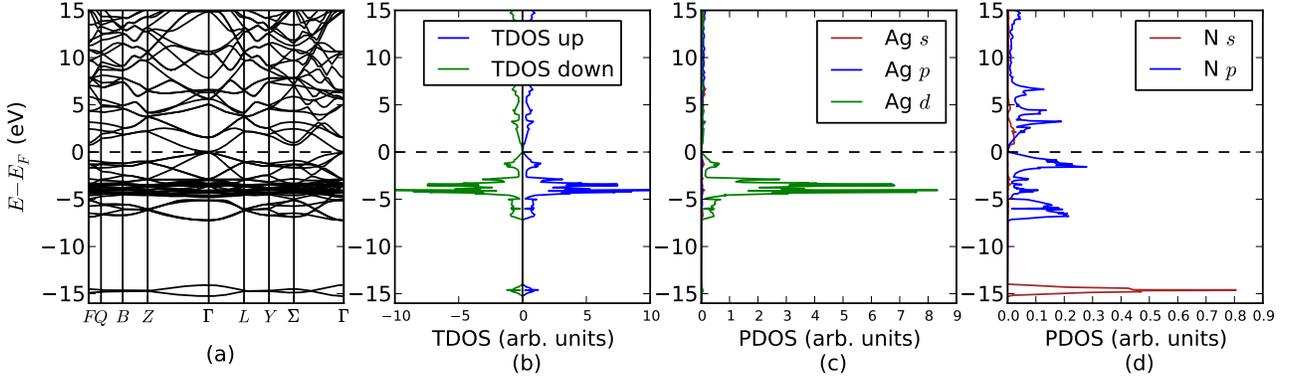}
\caption{\label{Ag3N1_RhF_3_electronic_structure}(Color online.) DFT calculated electronic structure for Ag$_3$N in the RhF$_3$ structure:
\textbf{(a)} band structure along the high-symmetry $\mathbf{k}$-points which are labeled according to Ref. [\onlinecite{Bradley}]. Their coordinates w.r.t. the reciprocal lattice basis vectors are: $F(0.5, 0.5, 0.0)$, $Q(0.375, 0.625, 0.0)$, $B(0.5, 0.75, 0.25)$, $Z(0.5, 0.5, 0.5)$, $\Gamma(0.0,  0.0, 0.0)$, $L(0.0, 0.5, 0.0)$, $Y(0.25, 0.5, -.25)$, $\Sigma(0.0, 0.5, -.5)$;
 \textbf{(b)} spin-projected total density of states (TDOS);
 \textbf{(c)} partial density of states (PDOS) of Ag($s, p, d$) orbitals in Ag$_3$N;
 and \textbf{(d)} PDOS of N($s, p$) orbitals in Ag$_3$N.}
\end{figure*}

\begin{figure*}
\includegraphics[width=1.0\textwidth]{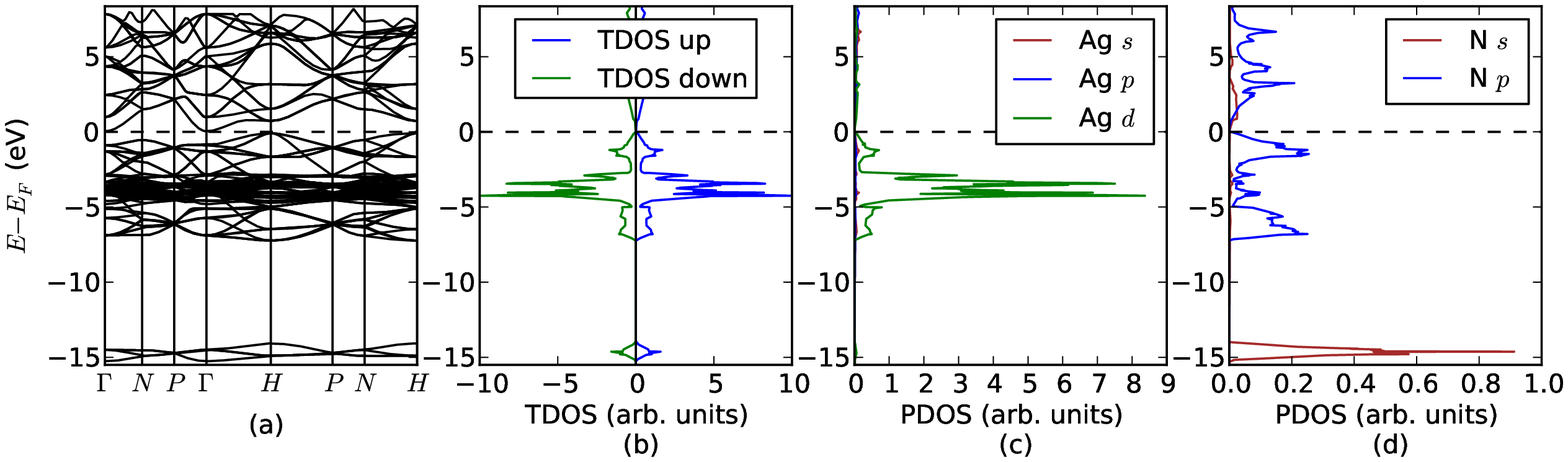}
\caption{\label{Ag3N1_D0_2_electronic_structure}(Color online.) DFT calculated electronic structure for Ag$_3$N in the D0$_2$ structure:
 \textbf{(a)} band structure along the high-symmetry $\mathbf{k}$-points which are labeled according to Ref. [\onlinecite{Bradley}]. Their coordinates w.r.t. the reciprocal lattice basis vectors are:  $\Gamma(0.0, 0.0, 0.0)$, $N(0.0, 0.0, 0.5)$, $P(0.25, 0.25, 0.25)$, $H(0.5, -.5, 0.5)$;
 \textbf{(b)} spin-projected total density of states (TDOS);
 \textbf{(c)} partial density of states (PDOS) of Ag($s, p, d$) orbitals in Ag$_3$N;
 and \textbf{(d)} PDOS of N($s, p$) orbitals in Ag$_3$N.}
\end{figure*}

\begin{figure*}
\includegraphics[width=1.0\textwidth]{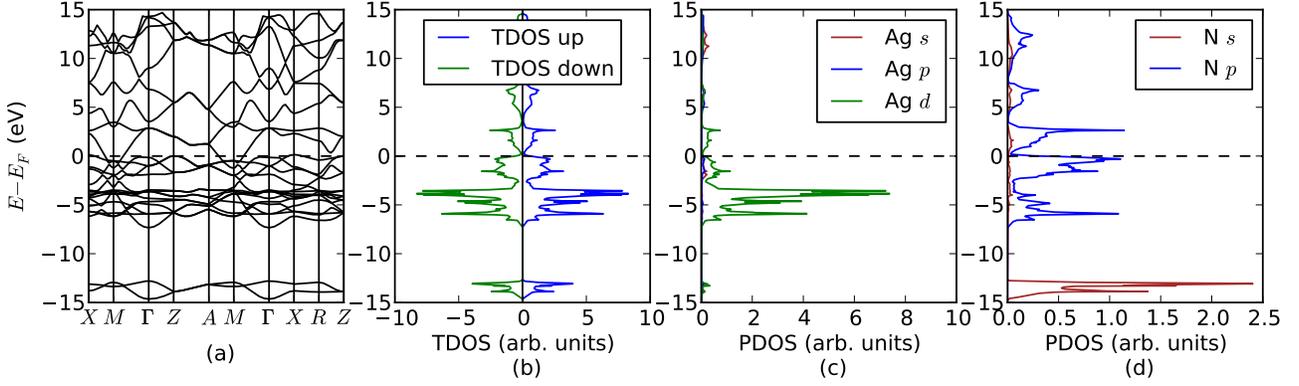}
\caption{\label{Ag1N1_B17_electronic_structure}(Color online.) DFT calculated electronic structure for AgN in the B17 structure:
 \textbf{(a)} band structure along the high-symmetry $\mathbf{k}$-points which are labeled according to Ref. [\onlinecite{Bradley}]. Their coordinates w.r.t. the reciprocal lattice basis vectors are: $X (0.0, 0.5, 0.0)$, $M (0.5, 0.5, 0.0)$, $\Gamma (0.0, 0.0, 0.0)$, $Z (0.0, 0.0, 0.5)$, $A (0.5, 0.5, 0.5)$, $R (0.0, 0.5, 0.5)$;
 \textbf{(b)} spin-projected total density of states (TDOS);
 \textbf{(c)} partial density of states (PDOS) of Ag($s, p, d$) orbitals in AgN;
 and \textbf{(d)} PDOS of N($s, p$) orbitals in AgN.}
\end{figure*}

\begin{figure*}
\includegraphics[width=1.0\textwidth]{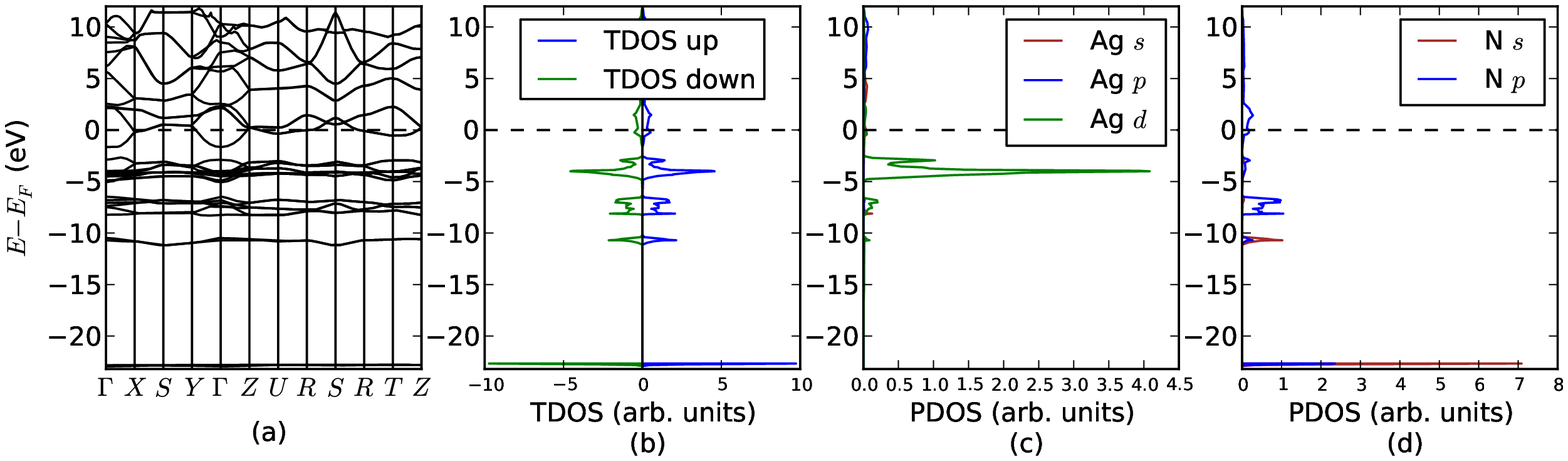}
\caption{\label{Ag1N2_C18_electronic_structure}(Color online.) DFT calculated electronic structure for AgN$_2$ in the C18 structure:
 \textbf{(a)} band structure along the high-symmetry $\mathbf{k}$-points which are labeled according to Ref. [\onlinecite{Bradley}]. Their coordinates w.r.t. the reciprocal lattice basis vectors are: $\Gamma( 0.0, 0.0, 0.0)$, $X( 0.0, 0.5, 0.0)$, $S( -.5, 0.5, 0.0)$, $Y( -.5, 0.0, 0.0)$, $Z( 0.0, 0.0, 0.5)$, $U( 0.0, 0.5, 0.5)$, $R( -.5, 0.5, 0.5)$, $T( -.5, 0.0, 0.5)$;
 \textbf{(b)} spin-projected total density of states (TDOS);
 \textbf{(c)} partial density of states (PDOS) of Ag($s, p, d$) orbitals in AgN$_2$;
 and \textbf{(d)} PDOS of N($s, p$) orbitals in AgN$_2$.}
\end{figure*}

The DFT(GGA) calculated band diagrams (i.e. $\epsilon_{i}^{\sigma}(\mathbf{k})$ curves) and spin-projected total and orbital resolved (i.e. partial) density of states (DOS) of the most stable phases: D0$_9$, RhF$_3$, D0$_2$, B17, and C18 are presented in Figs. \ref{Ag3N1_D0_9_electronic_structure}, \ref{Ag3N1_RhF_3_electronic_structure}, \ref{Ag3N1_D0_2_electronic_structure}, \ref{Ag1N1_B17_electronic_structure} and \ref{Ag1N2_C18_electronic_structure}, respectively. Spin-projected total density of states (TDOS) are shown in sub-figure (b) in each case. In all the six considered cases, electrons occupy the spin-up and spin-down bands equally, resulting in zero spin-polarization density of states: $\zeta(\epsilon) = n_{\uparrow}(\epsilon) - n_{\downarrow}(\epsilon)$. Thus, it is sufficient only to display spin-up (or spin-down) density of states (DOS) and spin-up (or spin-down) band diagrams. In order to investigate the details of the electronic structure of these phases, energy bands  are plotted along densely sampled high-symmetry string of neighboring $\mathbf{k}$-points. Moreover, to extract information about the orbital character of the bands, the Ag($s, p, d$) and N($s, p$) partial DOS are displayed at the same energy scale.
%
\begin{figure*}
\includegraphics[width=1.0\textwidth]{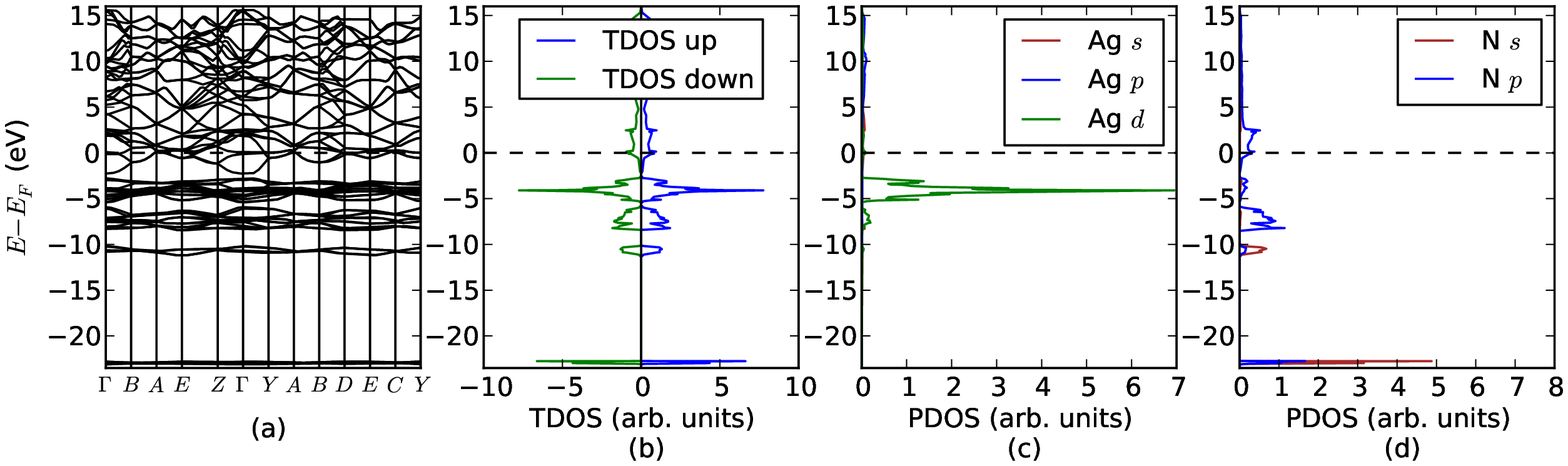}
\caption{\label{Ag1N2_CoSb2_electronic_structure}(Color online.) DFT calculated electronic structure for AgN$_2$ in the CoSb$_2$ structure:
 \textbf{(a)} band structure along the high-symmetry $\mathbf{k}$-points which are labeled according to Ref. [\onlinecite{Bradley}]. Their coordinates w.r.t. the reciprocal lattice basis vectors are: $\Gamma( 0.0, 0.0, 0.0)$, $B(-.5   ,  0.0  ,   0.0)$, $A(-.5   ,  0.5  ,   0.0)$, $E(-.5   ,  0.5  ,   0.5)$, $Z(0.0   ,  0.0  ,   0.5)$, $Y(0.0   ,  0.5  ,   0.0)$, $D(-.5   ,  0.0  ,   0.5)$ and $C(0.0   ,  0.5  ,   0.5)$;
 \textbf{(b)} spin-projected total density of states (TDOS);
 \textbf{(c)} partial density of states (PDOS) of Ag($s, p, d$) orbitals in AgN$_2$;
 and \textbf{(d)} PDOS of N($s, p$) orbitals in AgN$_2$.}
\end{figure*}	

Fig. \ref{Ag3N1_D0_9_electronic_structure}(a) shows the band structure $\epsilon_{i}^{\sigma}(\mathbf{k})$ of Ag$_3$N(D0$_9$). With its valence band maximum (VBM) at $(R,-0.086 \; eV)$ and its conduction band minimum (CBM) at $(\Gamma,0.049 \;eV)$, Ag$_3$N(D0$_9$) presents a semiconducting character with a narrow indirect band gap $E_g$ of $0.134 \; eV$. From sub-figures \ref{Ag3N1_D0_9_electronic_structure}(c) and (d), it is seen clearly that the Ag($d$)-N($p$) mixture in the region from $-7.286 \; eV$ to $-0.086 \; eV$ beneath $E_F$, with two peaks: a low density peak around $1.5 \; eV$ and a high density peak around $4.0 \; eV$ steaming mainly from the bands of silver $d$ electrons.

Our obtained PDOS, TDOS and band structure of Ag$_3$N(D0$_9$) agree qualitatively well with Gordienko and Zhuravlev \cite{AgN_AgN2_Ag2N_Ag3N_2010_comp}; however, using LCAO method within GGA, the value of the indirect $E_g$ of Ag$_3$N(D0$_9$) they predicted is $0.25 \; eV$.

To the best of our knowledge, there is no experimentally reported $E_{g}$ value for Ag$_3$N. However, Tong \cite{Ag_Pd_Au_nitrides_PhD_thesis_2010_exp} prepared Ag$_{3+x}$N samples, and carried out XRD measurements to confirm the fcc symmetry of the prapared samples. Using a TB-LMTO code within LDA, Tong then calculated the band structure of Ag$_{3}$N and obtained an indirect energy gap of $1.35 \; eV$. Nevertheless, we could not figure out the positions of the N ions Tong's model.
 
It is a well known drawback of Kohn-Sham DFT-based calculations to underestimate the band gap. Thus the more demanding $GW$ calculations were carried out, and the obtained $E_{g}$ value will be presented in Sec. \ref{Optical Properties}. 

Calculated electronic properties of Ag$_3$N(D0$_2$) are displayed in Fig. \ref{Ag3N1_D0_2_electronic_structure}. sub-figure \ref{Ag3N1_D0_2_electronic_structure}(a) shows the energy bands $\epsilon_{i}^{\sigma}(\mathbf{k})$ of Ag$_3$N(D0$_2$). With its valence band maximum (VBM) at $(H,-0.091 \; eV)$ and its conduction band minimum (CBM) at $(\Gamma,0.039 \;eV)$, Ag$_3$N(D0$_9$) presents semiconducting character with a narrow indirect band gap $E_g$ of $0.130 \; eV$. From sub-figures \ref{Ag3N1_D0_2_electronic_structure}(c) and (d), one can notice clearly the Ag($d$)-N($p$) mixture in the region from $-7.249 \; eV$ to $-0.091 \; eV$ below $E_F$, with two peaks: a low density peak around $-1.3 \; eV$ steaming from an almost equal mixture of Ag($d$) and N($p$), and a high density peak around $-4.3 \; eV$ steming mainly from the bands of silver $d$ electrons plus a relatively very low contribution from the N($p$) states.

Fig. \ref{Ag3N1_RhF_3_electronic_structure} depicts the band diagram and DOS's of Ag$_3$N(RhF$_3$). In contrast to Ag$_3$N(D0$_9$) and Ag$_3$N(D0$_2$), sub-figure \ref{Ag3N1_RhF_3_electronic_structure}(a) shows that Ag$_3$N(RhF$_3$) is a semiconductor with a narrow \textit{direct} band gap of $0.129 \; eV$ of width located at $\Gamma$ point. The VBM is at $-0.089 \; eV$ and the CBM is at $0.040 \;eV$. From sub-figures \ref{Ag3N1_RhF_3_electronic_structure}(c) and (d), one can see the Ag($d$)-N($p$) mixture is in the region from $-7.286 \; eV$ to $-0.089 \; eV$ beneath $E_F$, with two peaks: a low density peak around $-1.366 \; eV$ steaming from an almost equal mixture of Ag($d$) and N($p$), and a high density peak around $-4.382 \; eV$ steaming mainly from the bands of silver $d$ electrons plus a relatively very low contribution from the N($p$) states.

The relationship between D0$_9$, D0$_2$ and RhF$_3$ structures manifests itself in many common features between the electronic structure of these three Ag$_3$N nitrides: (i) equal $E_g$ of $\sim 0.13 \; eV$; (ii) a deep bound band around $\thicksim -14.6 \; eV$ below $E_F$ consists mainly of the N$(2s)$ states; (iii) a broad valence band with $\sim 7.2 \; eV$ of width that comes mostly from the $4d$ electrons of Ag plus a very small contribution from N$(2p)$; and (iv) the relatively low TDOS of the conduction bands.

Energy bands $\epsilon_{i}^{\sigma}(\mathbf{k})$, total density of states (TDOS) and partial (orbital-resolved) density of states (PDOS) of AgN(B17) are  shown in Figs. \ref{Ag1N1_B17_electronic_structure}. It is clear that AgN(B17) would be a true metal at its equilibrium. The major contribution to the very low TDOS around Fermi energy $E_F$ comes from the $2p$ states of the N atoms as it is evident from sub-figure \ref{Ag1N1_B17_electronic_structure}(d). Beneath $E_F$ lies a band with $\sim 7.3 \; eV$ of width, in which the main contribution is due to the Ag($4d$) states plus a small contribution from the N($2p$) states. While the N($2s$) states dominate the deep lowest region around $13.5 \; eV$, the low density unoccupied bands stem mainly from the N($2p$) states. The Fermi surface crosses two partly occupied bands: a lower one in the $X$-$M$, $\Gamma$-$Z$-$A$ and $\Gamma$-$X$-$R$ directions, and a higher band in the $X$-$M$-$\Gamma$ and $M$-$A$ directions. Thus, $E_F$ is not a continuous surface contained entirely within the first BZ.

It may be worth mentioning here that AgN(B1) \cite{CuN_AgN_AuN_2007_comp,AgN_2008_comp} and AgN(B3) \cite{CuN_AgN_AuN_2007_comp,PdN_AgN_2007_comp} phases were also theoretically predicted to be metallic.

Although AgN$_2$(CoSb$_2$) is the most stable phase, but the difference in cohesive energy between AgN$_2$(CoSb$_2$) and AgN$_2$(C18) is less than $0.02 \; eV/\text{atom}$, and we decided to examine the electronic structure of both phases. With $E_F$ crossing the finite TDOS, Fig. \ref{Ag1N2_C18_electronic_structure} shows that AgN$_2$(C18) is metallic at $0 \; K$. The orbital resolved DOS's reveal that the major contribution to the low TDOS at $E_F$ comes from the N($2p$) states with tiny contributions from the $5s$, $4d$ and $3p$ states of Ag, respectively. As one can see from sub-figure \ref{Ag1N2_C18_electronic_structure}(a), the $E_F$ surface crosses the edges of the first Brillouin zone in the $Z$-$U$-$R$-$S$-$T$-$X$ and $T$-$Z$ directions.

The calculated electronic properties of AgN$_2$(CoSb$_2$) are displayed in Fig. \ref{Ag1N2_CoSb2_electronic_structure}. Band structure, TDOS and orbital resolved DOS's have almost the same features as the corresponding ones of AgN$_2$(C18). It may be worth to mention here that C1 phase of AgN$_2$ was also theoretically predicted to be metallic \cite{AgN_AgN2_Ag2N_Ag3N_2010_comp}.

Compared to the metallic AgN(B17), three new features of these 1:2 nitrides are evident: (i) Deep at $\sim -22.7 \; eV$ there is a highly-localized mixture of the N($s$)-N($p$) states. However, the variation in N($2s$) energy with respect to \textbf{k} is smaller than the variation of N($2p$) states, resulting in a narrower and higher PDOS. (ii) Below the band that is crossed by $E_F$ there are four bands separated by $\sim 11.4 \; eV$, $\sim 1.6 \; eV$, $\sim 0.38 \; eV$ and $\sim 0.28 \; eV$ energy gaps, respectively. (iii) The very tiny contribution of the N($p$) states to the N($2p$)-Ag($4d$) band.

A common feature of all the studied cases is that Ag($p$)-orbitals do not contribute significantly to the hybrid bands. Another common feature is the highly structured, intense and narrow series of peaks in the TDOS valance band corresponding to the superposition of N($2p$) and Ag($4d$) states. In their $\mathbf{k}$-space, Ag($4d$) energies show little variation with respect to $\mathbf{k}$; hence the Van Hove singularities-like sharp features.

To summarize, we have found that the most stable phases of AgN and AgN$_{2}$ are metallic, while those of Ag$_{3}$N are semiconductors. A close look at Fig. \ref{Ag1N1_B17_electronic_structure} up to Fig. \ref{Ag3N1_D0_9_electronic_structure} reveals that as the nitrogen to silver ratio increases from $x = 1$ to $x = 1/2$, the TDOS at $E_{F}$ decreases; and by arriving at $x = 1/3$ a gap opens. This finding agrees well with Gordienko and Zhuravlev \cite{AgN_AgN2_Ag2N_Ag3N_2010_comp}. Moreover, it may be worth mentioning here that such behavior was theoretically predicted to be true for copper nitrides as well \cite{Suleiman_PhD_arXiv2012_copper_nitrides_article,Cu3N_Cu4N_Cu8N_2007_comp}.
%
\subsection{\label{Optical Properties}Optical Properties}
\begin{figure*}	
\includegraphics[width=1.0\textwidth]{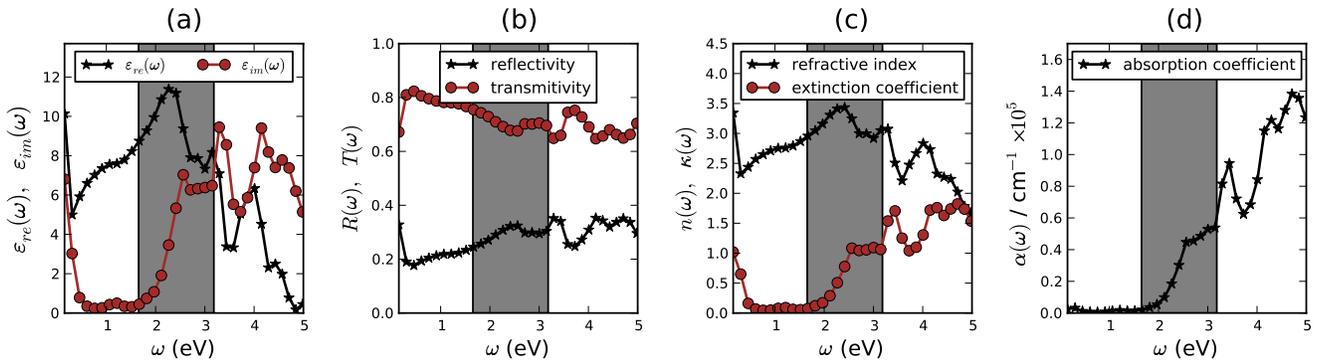}
\caption{\label{Ag3N1_D0_9_optical_constants}(Color online.) 
The $GW$ calculated frequency-dependent optical spectra of Ag$_3$N(D0$_9$):
 \textbf{(a)} the real $\varepsilon_{\text{re}}(\omega)$ and the imaginary $\varepsilon_{\text{im}}(\omega)$ parts of the dielectric function $\varepsilon_{\text{RPA}}(\omega)$;
 \textbf{(b)} reflectivity $R(\omega)$ and transmitivity $T(\omega)$;
 \textbf{(c)} refraction $n(\omega)$ and extinction $\kappa(\omega)$ coefficients; and 
 \textbf{(d)} absorption coefficient $\alpha(\omega)$.
 The shaded area highlights the optical region.}
\end{figure*}

Fig. \ref{Ag3N1_D0_9_optical_constants} depicts the $GW$ calculated real and imaginary parts of the frequency-dependent dielectric function $\varepsilon_{\text{RPA}}(\omega)$ of Ag$_3$N(D0$_9$) and the corresponding derived optical constants. The optical region\footnote{\textbf{\textit{Recall that the optical region (i.e. the visible spectrum) is about $(390 \sim 750) \; nm$ which corresponds to $(3.183 \sim 1.655) \; eV$.}}} is shaded in each sub-figure.

The real part $\varepsilon_{\text{re}}(\omega)$ (sub-figure \ref{Ag3N1_D0_9_optical_constants}(a)) shows an upward trend before $\sim 2.3 \; eV$, where it reaches its maximum value and generally decreases after that. The imaginary part $\varepsilon_{\text{im}}(\omega)$ (same sub-figure \ref{Ag3N1_D0_9_optical_constants}(a)) shows an upward trend before $\sim 1.0 \; eV$ and it has three main peaks located at $\sim 2.6 \; eV$ in the optical region, $\sim 3.3 \; eV$ at the right edge of the optical region, and at $\sim 4.1 \; eV$ in the UV range.

Calculated reflectivity $R(\omega)$ and transmitivity $T(\omega)$ are displayed in sub-figure \ref{Ag3N1_D0_9_optical_constants}(b). With $0.6 \leq R(\omega) \leq 0.8$, it is evident that Ag$_3$N(D0$_9$) is a good reflector, specially in the red and the infrared regions. In the visible range, the maximum transmitivity $T(\omega)$ is at $\sim 2.54 \; eV \equiv 489 \; \text{nm}$, which is at the blue-green edge.

sub-figure \ref{Ag3N1_D0_9_optical_constants}(c) depicts the calculated refraction $n(\omega)$ and extinction $\kappa(\omega)$ coefficients. As they should, these two spectra have, in general, the same qualitative frequency dependence as the real $\varepsilon_{\text{re}}(\omega)$ and the imaginary $\varepsilon_{\text{im}}(\omega)$ dielectric functions, respectively.

From the absorption coefficient $\alpha\left(\omega\right)$ spectrum (sub-figure \ref{Ag3N1_D0_9_optical_constants}(d)), it can be seen that Ag$_{3}$N(D0$_{9}$) starts absorbing photons with $\sim 0.9 \; eV$ energy. Hence, it is clear that $GW_{0}$ calculations give a band gap of $\sim 0.9 \; eV$, which is a significant improvement over the value obtained from DFT. The non-vanishing $\alpha\left(\omega\right)$ in the whole optical region agrees with the experiment, since it may explain the observed black color of the synthesized Ag$_{3}$N.

To the best of our knowledge, the present work is the first trial to theoretically investigate the optical properties of silver nitride. However, for more accurate optical characterization (e.g. more accurate positions and amplitudes of the characteristic peaks), electron-hole excitations should be calculated. This can be done by evaluating the two-body Green function $G_2$ on the basis of our obtained GW one-particle Green function $G$ and QP energies, then solving the so-called Bethe-Salpeter equation, the equation of motion of $G_2$ \cite{DFT_GW_BSE_Electron-hole_excitations_and_optical_spectra_from_first_principles_2000}.
%
\section{\label{Conclusions}Conclusions}
We have succesfully employed first-principles calculation methods to investigate the structural, stability, electronic and optical properties of Ag$_3$N, AgN and AgN$_2$. Within the accuracy of the employed methods, the obtained structural parameters, EOS, $B_{0}$, $B_{0}^{\prime}$ and electronic properties show good agreement with the few avialable previous calculations. On the other hand, our obtained results show, at least, partial agreement with three experimental facts: (i) the lattice parameter of Ag$_3$N(D0$_9$) is close to the experimentally reported one; (ii) the positive formation energies reveals the endothermic (unstable) nature of silver nitrides, and (iii) absorption spectrum explains its observed black color. Moreover, the present work may be considered as the first trial to theoretically investigate the optical properties of silver nitride. We hope that some of our obtained results will be confirmed in future experimentally and/or theoretically.
%
\section*{Acknowledgments}
All GW calculations and some DFT calculations were carried out using the infrastructure of the Centre for High Performance Computing (CHPC) in Cape Town. Suleiman would like to acknowledge the support he received from Wits, DAAD, AIMS, SUST and the ASESMA group. Many thanks to the Scottish red pen of Ross McIntosh!
%
\bibliography{v1_arXiv_silver_nitrides_article}
\end{document}

%% file: footed_results_table.tex
\begin{table*}
\tiny
\caption{\label{silver_nitrides_equilibrium_structural_properties}Calculated and experimental zero-pressure properties of the twenty studied phases of Ag$_{3}$N, AgN and AgN$_2$\textbf{:} Lattice constants ($a(\text{\AA})$, $b(\text{\AA})$, $c(\text{\AA})$, $\alpha(^{\circ})$ and $\beta(^{\circ})$), equilibrium atomic volume $V_{0}(\text{\AA}^{3}/$atom$)$, cohesive energy $E_{\text{coh}} (eV/$atom$)$, bulk modulus $B_{0} (GPa)$ and its pressure derivative $B_{0}^{\prime}$, and formation energy $E_{f}(eV/\text{atom})$. The presented data are of the current work (\textit{Pres.}), experimentally reported (\textit{Expt.}) and of previous calculations (\textit{\textit{Comp.}}).}
\resizebox{1.0\textwidth}{!}{
\begin{tabular}{lllllllllll}
\hline
\textbf{Structure}	&		& $a(\AA)$		   & $b(\AA)$			& $c(\AA)$		  & $\alpha(^{\circ})$ or $\beta(^{\circ})$		& $V_{0} (\AA^{3}/$atom$)$   & $E_{\text{coh}}(eV/\text{atom})$	& $B_{0}(\text{GPa})$		& $B_{0}^{\prime}$ & $E_{f}(eV/\text{atom})$\\
\hline \hline 
\multicolumn{10}{c}{\textbf{Ag}} \\
\hline 
\multirow{4}{*}{\textbf{A1}}       & \textit{Pres.} & $4.164$ &--                         &--                         &--                         & $18.06$ & $-2.484$ & $88.188$ & $5.793$	&\\
       & \textit{Expt.}   & ($4.08570 \pm 0.00018$)\footnotemark[1] &--                         &--                         &--                      &   &  $-2.95$\footnotemark[2]   & $100.7$\footnotemark[2], $101$\footnotemark[3]   & $6.12$\footnotemark[4]	&\\
      		&	\multirow{2}{*}{\textit{Comp.}} &  \multirow{2}{*}{$4.01$\footnotemark[6]} &\multirow{2}{*}{--}       &\multirow{2}{*}{--}       &\multirow{2}{*}{--}  
& & $-3.59$\footnotemark[7], $-2.66$\footnotemark[8],   & 		
$142$\footnotemark[6] & $5.00$\footnotemark[12], $5.70$\footnotemark[10],                                 
                               		& \\     
        		& 	&   &   &    & &     &  $-2.67$\footnotemark[9] &          & $5.97$\footnotemark[11] 	& \\

\hline 
		\multicolumn{10}{c}{\textbf{Ag$_3$N}} \\
\hline 
\multirow{1}{*}{\textbf{D0$_3$}}        & \textit{Pres.}	& $6.322$ &--                         &--                         &--                         & $15.79$ & $-2.055$ & $98.356$ & $5.457$	& $1.107$ \\

\hline 
\multirow{1}{*}{\textbf{A15}}           & \textit{Pres.}	& $5.065$ &--                         &--                         &--                         & $16.24$ & $-1.976$ & $92.280$ & $5.470$	& $1.186$ \\

\hline 
\multirow{2}{*}{\textbf{D0$_9$}}        & \textit{Pres.}	& $4.328$ &--                         &--                         &--                         & $20.27$ & $-2.513$ & $71.980$ & $5.386$	& $0.649$ \\
                                        & \textit{Comp.}	&   $3.995$\footnotemark[17], $4.169$\footnotemark[18], $4.292$\footnotemark[19] &--                         &--                         &--                         &                     &                     & $95.7$\footnotemark[18], $87.1$\footnotemark[19]   &  	& \\

\hline 
\multirow{1}{*}{\textbf{L1$_2$}}        & \textit{Pres.}	& $3.972$ &--       &--                         &--                         & $15.67$ & $-2.081$ & $100.743$ & $5.530$	& $1.081$ \\

\hline 
\multirow{1}{*}{\textbf{D0$_2$}}        & \textit{Pres.}	& $8.662$ &--                         &--                         &--                         & $20.31$ & $-2.514$ & $72.230$ & $5.335$	& $0.648$ \\

\hline 
\multirow{1}{*}{\textbf{$\epsilon$-Fe$_3$N}}& \textit{Pres.}	& $5.967$ &--     & $5.560$ &--         & $21.43$ & $-2.469$ & $64.737$ & $2.335$	& $0.692$ \\

\hline 
\multirow{1}{*}{\textbf{RhF$_3$}}       & \textit{Pres.}	& $6.126$ &--   &--      & $\alpha=59.989$ & $20.31$ & $-2.514$ & $72.237$ & $5.396$	& $0.648$ \\
\hline 
\textbf{fcc}\footnotemark[20]		&\textbf{Expt.}       & $4.369$\footnotemark[21], $4.29$\footnotemark[22], $4.378$\footnotemark[24] &--                         &--                         &      &      &        &          &  	&$2.587 \pm 0.364$\footnotemark[23]\\ 

\hline 
\multicolumn{10}{c}{\textbf{AgN}} \\
\hline 
\multirow{3}{*}{\textbf{B1}}   		& \textit{Pres.}	& $4.617$ & --                        & --	                   & --                        & $12.30$ & $-2.253$ & $147.600$	& $5.145$ 		& $1.587$ \\
              		& \textit{Comp.} & $4.57$\footnotemark[17], $4.506$\footnotemark[18], $4.619$\footnotemark[19]	& --			       & --			   & -- 		       &			   &		       & $219.2$\footnotemark[18], $162.3$\footnotemark[19]	& $4.653$\footnotemark[16]  & \\
                           & 	&  $4.476$\footnotemark[15], $4.606$\footnotemark[16]    &--                         & --           &--          &        &         & $197.18$\footnotemark[15], $147.40$\footnotemark[16]  & $4.883$\footnotemark[15] 	&\\    

\hline 
\multirow{3}{*}{\textbf{B2}}   		& \textit{Pres.}	& $2.873$ & --	                & --	                    &--                       & $11.86$ & $-2.021$ & $146.157$ & $5.260$		 	& $1.819$ \\
           	& \multirow{2}{*}{\textit{Comp.}} & $2.833$\footnotemark[17], $2.806$\footnotemark[18], $2.876$\footnotemark[19]	& --	                & --	                    &--                       &                     &                     & $138.96$\footnotemark[16] & $4.823$\footnotemark[16]	&\\     
       		&	& $2.78$\footnotemark[15], $2.87$\footnotemark[16]   &--                         &                     &--                       &                     &                     & $204.10$\footnotemark[15] & $5.451 
$\footnotemark[15]  	&\\     

\hline 
\multirow{3}{*}{\textbf{B3}}   		& \textit{Pres.}	&  $4.950$ & --	                & --	                    &--                       & $15.16$ & $-2.122$ & $109.639$ & $5.210$	& $1.718$ \\
          	& \multirow{2}{*}{\textit{Comp.}} & $4.88$\footnotemark[17], $4.816$\footnotemark[18], $4.946$\footnotemark[19]	& --	                & --	                    &--                       &                     &                     & $100.11$\footnotemark[16]  & $5.825$\footnotemark[16] 	&\\     
                           & 	&  $4.79$\footnotemark[15], $4.94$\footnotemark[16]    &--                         & --           &--          &        &         & $151.05$\footnotemark[15]  & $4.542$\footnotemark[15] 	&\\     

\hline 
\multirow{1}{*}{\textbf{B8$_{1}$}}   	& \textit{Pres.}	&  $3.544$ &--  		        & $4.929$ &--                       & $13.40$ & $-1.996$ & $130.485$	&	$5.240$	& $1.844$ \\

\hline 
\multirow{1}{*}{\textbf{B$_{\text{k}}$}}& \textit{Pres.}	&  $3.521$ &--                         & $9.368$                    &--                       & $25.15$ & $-1.891$ & $57.077$ & $5.110$	& $1.949$ \\

\hline 
\multirow{1}{*}{\textbf{B$_{\text{h}}$}}& \textit{Pres.}	&  $3.096$ &--                         & $3.023$                    &--                       & $12.55$ & $-2.121$ & $141.385$ & $5.285$	& $1.719$ \\

\hline 
\multirow{2}{*}{\textbf{B4}}   		& \textit{Pres.}	&  $3.501$ &--                         & $5.734$ &--                       & $15.22$ & $-2.113$ & $105.992$ & $5.467$ 	& $1.727$ \\
                               		& \textit{Comp.}	&  $3.41$\footnotemark[15], $3.54$\footnotemark[16] &--                         &  $5.52$\footnotemark[15], $5.69$\footnotemark[16]  &--                       &                     &                     & $143.68$\footnotemark[15], $110.12$\footnotemark[16] & $4.82$\footnotemark[15], $4.663$\footnotemark[16]	&\\     

\hline 
\multirow{1}{*}{\textbf{B17}}   	& \textit{Pres.}	& Pres.	$3.158$ &--                         & $5.560$ &--                       & $13.86$ & $-2.517$ & $132.556$ & $5.185$ 	& $1.323$ \\

\hline 
\multirow{1}{*}{\textbf{B24}}   	& \textit{Pres.}	& $4.337$ & $4.601$ & $5.091$ &--                       & $12.70$ & $-2.202$ & $138.704$ & $5.132$ 	& $1.638$ \\
\hline 
		\multicolumn{10}{c}{\textbf{AgN$_2$}} \\
\hline 
\multirow{3}{*}{\textbf{C1}}            & \textit{Pres.}	& $5.157$ &--                         &--                         &--                         & $11.43$ & $-1.959$ & $164.844$ & $4.996$	 	& $2.333$ \\
                      & \multirow{2}{*}{\textit{Comp.}}	& $5.124$\footnotemark[17], $5.055$\footnotemark[18], $5.172$\footnotemark[19]  &--                         &--                         &--                         &                     &                     & $181.3$\footnotemark[18], $164.5$\footnotemark[19]    &  	& \\
                     &  & $5.013$\footnotemark[13], $5.141$\footnotemark[14]                     &--                         &--                         &--                         &                     &                     &  $215$\footnotemark[13], $164$\footnotemark[14]  &  	&\\

\hline 
\multirow{1}{*}{\textbf{C2}}            & \textit{Pres.}	& $5.617$ &--                         &--                         &--                         & $14.77$ & $-3.626$ & $ 30.058$         & $6.894$ 	& $0.666$ \\

\hline 
\multirow{1}{*}{\textbf{C18}}           & \textit{Pres.}	& $3.440$ & $4.513$ & $5.508$ &--                         & $14.25$ & $-3.680$ & $35.878$ & $7.269$ 	& $0.612$ \\

\hline 
\multirow{1}{*}{\textbf{CoSb$_2$}}      & \textit{Pres.}	& $5.976$ & $5.651$	& $10.261$ & $\beta = 151.225$        & $13.90$ & $-3.699$ & $35.117$ & $7.822$  & $0.593$ \\

\hline \hline   
\end{tabular}
}       

\footnotetext[1]{Ref. \cite{Jerry_1974}. This is an average of 56 experimental values, at $20 ^{\circ} C$.}
\footnotetext[2]{Ref. \cite{Kittel}. Cohesive energies are given at $0 \; K$ and $1 \text{ atm} = 0.00010 \;  \text{GPa}$; while bulk mudulii are given at room temperature.}
\footnotetext[3]{Ref. (25) in \onlinecite{B_prime_1997_theory_comp_n_exp}: at room temperature.}
\footnotetext[4]{See Refs. (8)--(11) in \onlinecite{B_prime_1997_theory_comp_n_exp}.}

\footnotetext[6]{Ref. \cite{elemental_metals_1996_comp}. using the full-potential linearized augmented plane waves (LAPW͒) method within LDA.}
\footnotetext[7]{Ref. \cite{elemental_metals_2008_comp}: using projector augmented wave (PAW) method within LDA.}
\footnotetext[8]{Ref. \cite{elemental_metals_2008_comp}: using projector augmented wave (PAW) method within GGA(PW91).}

\footnotetext[9]{Ref. \cite{elemental_metals_2008_comp}: using projector augmented wave (PAW) method within GGA(PBE).}
\footnotetext[10]{Ref. \cite{B_prime_1997_theory_comp_n_exp}: using a semiempirical estimate based on the calculation of the slope of the shock velocity \textit{vs.} particle velocity curves obtained from the dynamic high-pressure experiments. The given values are estimated at $\sim 298 \; K$.}
\footnotetext[11]{Ref. \cite{B_prime_1997_theory_comp_n_exp}: using a semiempirical method in which the experimental static $P-V$ data are fitted to an EOS form where $B_{0}$ and $B_{0}^{\prime}$ are adjustable parameters. The given values are estimated at $\sim 298 \; K$.}
\footnotetext[12]{Ref. \cite{B_prime_1997_theory_comp_n_exp}: using the so-called method of transition metal pseudopotential theory; a modified form of a method proposed by Wills and Harrison to represent the effective interatomic interaction.}

\footnotetext[13]{Ref. [\onlinecite{AgN2_AuN2_PtN2_2005_comp}]: using the full-potential linearized augmented plane waves (LAPW͒) method within LDA.}
\footnotetext[14]{Ref. [\onlinecite{AgN2_AuN2_PtN2_2005_comp}]: using the full-potential linearized augmented plane waves (LAPW͒) method within GGA.}
\footnotetext[15]{Ref. [\onlinecite{CuN_AgN_AuN_2007_comp}]: using full-potential (linearized) augmented plane waves plus local orbitals (FP-LAPW+lo) method within LDA.}
\footnotetext[16]{Ref. [\onlinecite{CuN_AgN_AuN_2007_comp}]: using using full-potential (linearized) augmented plane waves plus local orbitals (FP-LAPW+lo) method within GGA(PBE).}
\footnotetext[17]{Ref. [\onlinecite{AgN_AgN2_Ag2N_Ag3N_2010_comp}]: using pseudopotential (PP) method within LDA.}
\footnotetext[18]{Ref. [\onlinecite{AgN_AgN2_Ag2N_Ag3N_2010_comp}]: using linear combinations of atomic orbitals (LCAO) method within LDA. $B_0$'s are calculated from elastic constants.}
\footnotetext[19]{Ref. [\onlinecite{AgN_AgN2_Ag2N_Ag3N_2010_comp}]: using linear combinations of atomic orbitals (LCAO) method within GGA. $B_0$'s are calculated from elastic constants.}
\footnotetext[20]{This is the face centered cubic (fcc) structure with $Z = 4/3$ (i.e. 4 Ag atoms in the unit cell) suggested by Hahn and Gilbert according to some measurements (Ref. \onlinecite{German_Ag3N_structure_1949_exp}).}
\footnotetext[21]{Ref. \onlinecite{German_Ag3N_structure_1949_exp}.}
\footnotetext[22]{Ref. \onlinecite{Ag_Pd_Au_nitrides_PhD_thesis_2010_exp}.}
\footnotetext[23]{This is the average of the experimental values:
 $(+314.4 \mp 2.5) \; \text{kJ/mol}$  \cite{Ag3N_1991_exp} = $(3.25853 \pm 0.02591) \; eV/\text{atom}$,
 $+199 \; \text{kJ/mol}$ \cite{Juza_n_Hahn_1940} = $2.062 \; eV/\text{atom}$,
 $+255 \; \text{kJ/mol}$ \cite{German_Ag3N_structure_1949_exp} = $2.643 \; eV/\text{atom}$, and
 $+230 \; \text{kJ/mol}$ \cite{Anderson_n_Parlee_1970_exp} = $2.384 \; eV/\text{atom}$.
 We used the conversion relation: $1 \; \text{eV/atom} = 96.521 \; \text{kJ/mol}$ or equivalently $1 \; \text{kJ/mol} = 0.010364 \; \text{eV/atom}$.}
\footnotetext[24]{Ref. \onlinecite{Ag3N_structure_theory_1982}.}

\end{table*}